\documentclass{sig-alternate}

\newcommand{\ignore}[1]{}
\usepackage{fancyhdr}
\usepackage[normalem]{ulem}
\usepackage[hyphens]{url}
\usepackage{hyperref}

\usepackage{subfig}
\usepackage{multirow}
\usepackage{array}
\usepackage{xcolor,colortbl}
\usepackage[justification=centering]{caption}
\usepackage{siunitx}
\usepackage{enumitem}
\usepackage{subscript}
\usepackage{gensymb}

\newcommand{\upperRomannumeral}[1]{\uppercase\expandafter{\romannumeral#1}}

\fancypagestyle{firstpage}{
  \fancyhf{}
  \pagenumbering{arabic}
}  

\title{HAPPY: Hybrid Address-based Page Policy in DRAMs}
\subtitle{}
\author{Mohsen Ghasempour$^\dagger$, Aamer Jaleel$^{\star}$, Jim Garside$^{\dagger}$ and Mikel Luj{\'a}n$^{\dagger}$\\
School of Computer Science, University of Manchester$^{\dagger}$ \\
  NVidia Research$^{\star}$}

\begin{document}
\maketitle
\thispagestyle{firstpage}
\pagestyle{plain}

%%%%%% -- PAPER CONTENT STARTS-- %%%%%%%%

\begin{abstract}

Memory controllers have used static page closure policies to decide whether a row should be left open, \textit{open-page policy}, or closed immediately, \textit{close-page policy}, after the row has been accessed.  The appropriate choice for a particular access can reduce the average memory latency. However, since application access patterns change at run time, static page policies cannot guarantee to deliver optimum execution time. Hybrid page policies have been investigated as a means of covering these dynamic scenarios and are now implemented in state-of-the-art processors. Hybrid page policies switch between open-page and close-page policies while the application is running, by monitoring the access pattern of row hits/conflicts and predicting future behavior. Unfortunately, as the size of DRAM memory increases, fine-grain tracking and analysis of memory access patterns does not remain practical. 

We propose a compact memory address-based encoding technique which can improve or maintain the performance of DRAMs page closure predictors while reducing the hardware overhead in comparison with state-of-the-art techniques. As a case study, we integrate our technique, HAPPY, with a state-of-the-art monitor -- the Intel-adaptive open-page policy predictor employed by the Intel Xeon X5650 -- and a traditional Hybrid page policy. We evaluate them across 70 memory intensive workload mixes consisting of single-thread and multi-thread applications. The experimental results show that using the HAPPY encoding applied to the Intel-adaptive page closure policy can reduce the hardware overhead by 5$\times$ for the evaluated 64~GB memory (up to 40$\times$ for a 512~GB memory) while maintaining the prediction accuracy.

% while improving the prediction accuracy of Intel-adaptive up to 10\%. 

%Similarly, the results also show a 182,000$\times$ reduction in cost of implementation when compared with existing Hybrid techniques for the 64~GB memory (up to 600,000$\times$ for 256~GB) while improving the performance of hybrid predictor up to 15\%. 
%To sump up, our technique achieves similar (or slightly better) performance as existing high performance industry predictors as well as published one at much less hardware overhead.

\end{abstract}

\section{Introduction}

The performance of DRAM is sensitive to the memory access pattern of the running applications.  
%Access to a DRAM device can be categorized in three ways: page-empty, page-hit and page-miss. If a memory request goes to the bank which has no open row the access called `page-empty'. In this situation an activation command is required to open the desired row. If a memory request goes to the row which is already open then it is a `page-hit' and the request can be serviced fairly cheaply.  Finally, if a memory request goes to a different row from the open one in a bank it is a `page-miss' which imposes an extra latency on the memory system; the open row must be closed first  before the desired row can be opened.
Traditionally DRAM controllers have used a static row-buffer access policy, either \textit{open-page} or \textit{close-page}, to decide whether a row should be left open or closed immediately after their access. For workloads with high locality of accesses open-page works best since the target row is already open and multiple accesses to that row can be serviced with one activation. However, for workloads with more random memory accesses, close-page is a better option. In this case a row will be closed immediately after a memory access so the next memory request within the same bank does not need to wait for the precharge process of the open row. Moreover, neither the open-page nor close-page policy can deliver the `best' execution time for all the workloads due to the dynamic nature of memory accesses. In this situation a \textit{hybrid-page} policy, which is a mixture of open-page and close-page, is more desirable.

Different techniques have been proposed in the literature to select between open-page and close-page in DRAM memory controllers. \textit{Access-based techniques} are those that monitor and keep a history of the row hits and row misses at different granularity in DRAMs to make a prediction of the future page closure policy.  On the other hand, \textit{time-based techniques} focus on predicting the optimum time that a row can be left open. In general, these techniques rely on predictors that monitor the number of accesses per row, the number of row hits or row misses, the time between hits or misses, etc.\ to predict the open-page or close-page for each row in DRAM. Intel included in the Xeon X5650 two time-based techniques.

As the size of DRAM is increasing with Data Analytic applications, having a fine-grain prediction and monitoring scheme is inefficient and not scalable. On the other hand going toward the coarse-grain schemes reduces the accuracy of the prediction. A key challenge for page-closure techniques is to balance the hardware overhead and the prediction accuracy. 

%The scalability issue of the existing page closure prediction algorithms becomes more important when considering the emergence of new memory technologies, like the Hybrid Memory Cube (HMC). HMC, introduced by Micron, is a 3D stacked multibank DRAM that provides 15$\times$ more bandwidth than conventional DDR3 modules and requires 70\% less energy and 90\% less space than existing modern memory technologies \cite{website:Micron2}. It has more data banks than traditional DRAMs and its unique 3D structure makes it extremely scalable in comparison with normal DRAM systems. 

%On the other hand, 

The trend towards keeping entire databases in DRAMs, such as RAMCloud (e.g.\ 64~TB of DRAMs) \cite{website:RAMCloud} or Facebook using 150~TB of DRAMs with memcache \cite{ongaro2011fast}, turns the scalability issue into a critical problem for future DRAM systems. 
%As new memory technology is going toward scalable design the non-scalable prediction techniques are not going to remain practical. 

Our contribution is a scalable and compact memory address-based encoding technique, called \textit{HAPPY}, that can be employed in DRAM memory controllers. HAPPY is an efficient encoding  that reduces the cost of implementation of existing page closure techniques while maintaining the prediction accuracy of the original implementation. As case studies, we show how to integrate HAPPY with a state-of-the-art implementation -- the Intel-adaptive open-page policy employed by Intel -- and with a traditional hybrid-page. We evaluate HAPPY and existing techniques across 70 memory intensive workload mixes consisting of single-thread and multi-thread applications. The experimental results show that using the HAPPY memory address-based encoding applied to the Intel-adaptive page policy can reduce the hardware cost of implementation by 5$\times$ for the evaluated 64~GB memory while maintaining the prediction accuracy. In other words, we can achieve similar, or better, performance as existing high performance industry and academic techniques while requiring less hardware overhead.

\section{Background and Motivation}

\begin{figure}[!htb]
	\centering
	%\begin{subfigure}
	\subfloat[DRAM Device]{\includegraphics[scale=0.2]{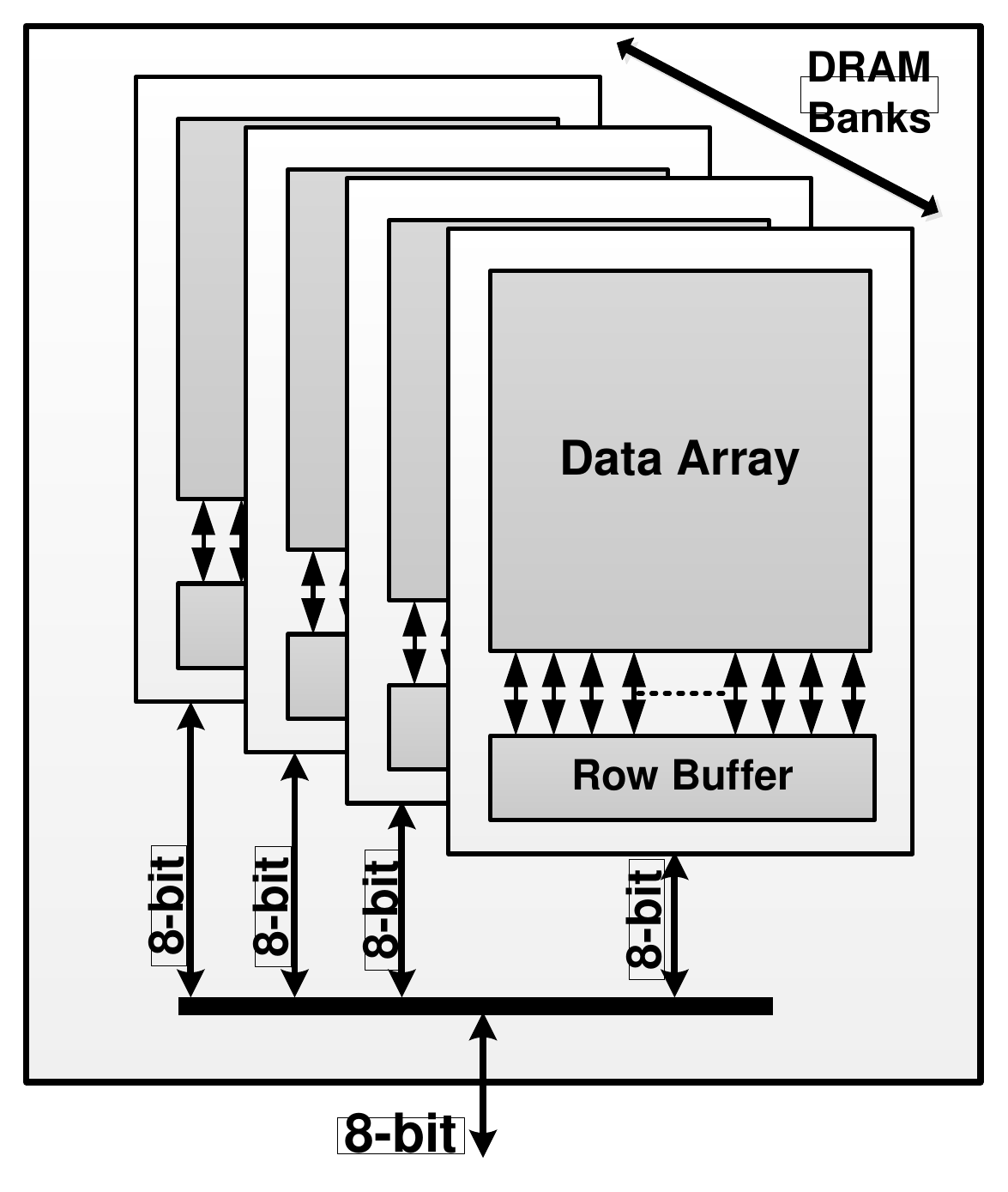}\label{fig:dram_device}} 
	\subfloat[DRAM Rank]{\includegraphics[scale=0.3]{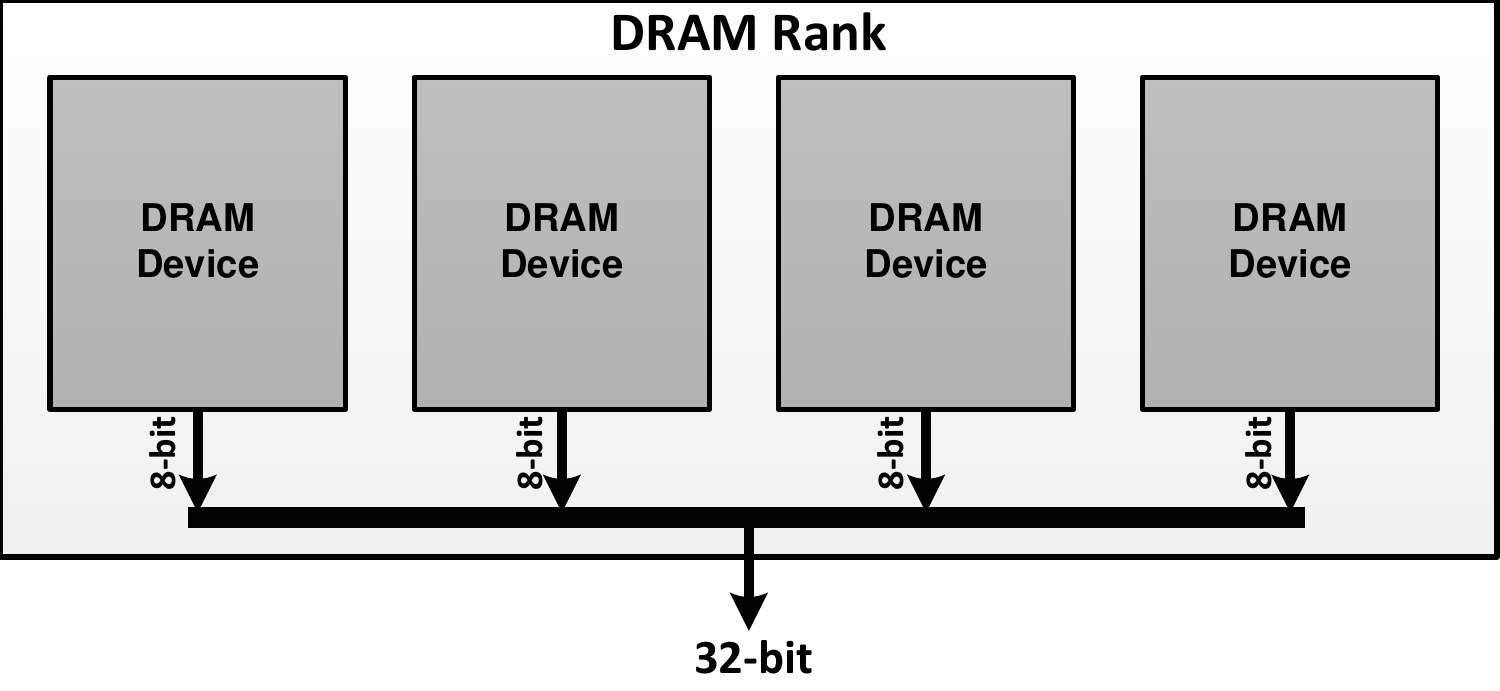}\label{fig:dram_rank}} 		

	\caption{DRAM Structure.}	
	\label{fig:DRAM_Organisation}
	%\end{subfigure}
\end{figure}

{\bf DRAM Structure:} Figure~\ref{fig:DRAM_Organisation} presents a high-level structure of a typical DRAM organization. A DRAM device (Figure~\ref{fig:dram_device}) consists of multiple banks, each of which includes a data array and a sense amplifier or \textit{row buffer}. The data array is a matrix of rows and columns comprising the storage cells. The basic operation of DRAMs requires that to access a specific cell within a bank the entire row (e.g.\ {1~KB} data) has to be moved into the row buffer. Then, read or write operations can be performed on the data stored in the row buffer. Although the banks within a DRAM device can be accessed in parallel, since they share the communication bus only one bank at the time can transfer data out of the DRAM device. Each DRAM device typically supports read/write operations of 4-16 bits per memory request depending on the DRAM model. To support the required bandwidth, multiple DRAM devices work in parallel within a Rank (Figure~\ref{fig:dram_rank}). In modern DRAMs, 64-bit data can be read/written per cycle and typically a burst of size 4 or 8 is supported by these modules to fill a full cache-line \cite{jacob2010memory,itoh2001vlsi}.

{\bf DRAM Basic Operation:} To perform a read or write operation, the target row has to be opened first using an \textit{activation} command which transfers a row to the row buffer, imposing a delay t\textsubscript{RCD}.  When the row is in the row buffer, a read or write command can be issued with a delay t\textsubscript{CL}. Considering the internal structure of DRAMs, only one row can be processed at a time. Thus, to access to a different row (within the same bank), the open row has to be closed first using the \textit{precharge} command with a delay t\textsubscript{RP}. This command prepares the row buffer to accept the new row. Considering the basic operation of DRAMs, each memory request can be classified into one of the following three categories depending on the status of the bank to be accessed: \textit{page-hit}, \textit{page-miss} or \textit{page-empty}.

A page-hit is defined as a read/write operation to an open row within a bank. In this situation there is no need to use an activation command and the memory request can be serviced immediately. A page-miss is defined as a read/write operation to a different row than the open row within a bank. In this situation the open row must first be closed before accessing the second row. Finally, a page-empty is defined as a read/write command to a bank that has no open row in the row buffer. In this case an activation command is required to open the target row. Page-misses are the most expensive memory request while the page-hits are the cheapest to service.  Page-empties are cheaper than the page conflicts but more expensive than the page-hits. 

{\bf DRAM Static Page Closure Policies:} DRAM memory controllers have a page closure policy, to alleviate the effect of page-misses on the memory system's performance. The traditional schemes are the open-page and the close-page policy. A DRAM that uses the open-page policy would leave the last accessed row open in the row buffer to eliminate the activation cost of the next memory request to the same row. A DRAM that uses the close-page policy would close the row immediately after it has been accessed to eliminate the possibility of getting a page-miss for the next memory request \cite{keeth2008dram}. In general, the open-page policy is more beneficial for the systems with high access locality whereas the close-page policy is more appropriate for systems with high entropy memory access. Table~\ref{table:Cost_of_different_page_closure_policies} presents the timing cost of page-hits and page-misses when using the static page closure policies.

\begin{figure*}[!htb]
\centering
\includegraphics[scale=0.27]{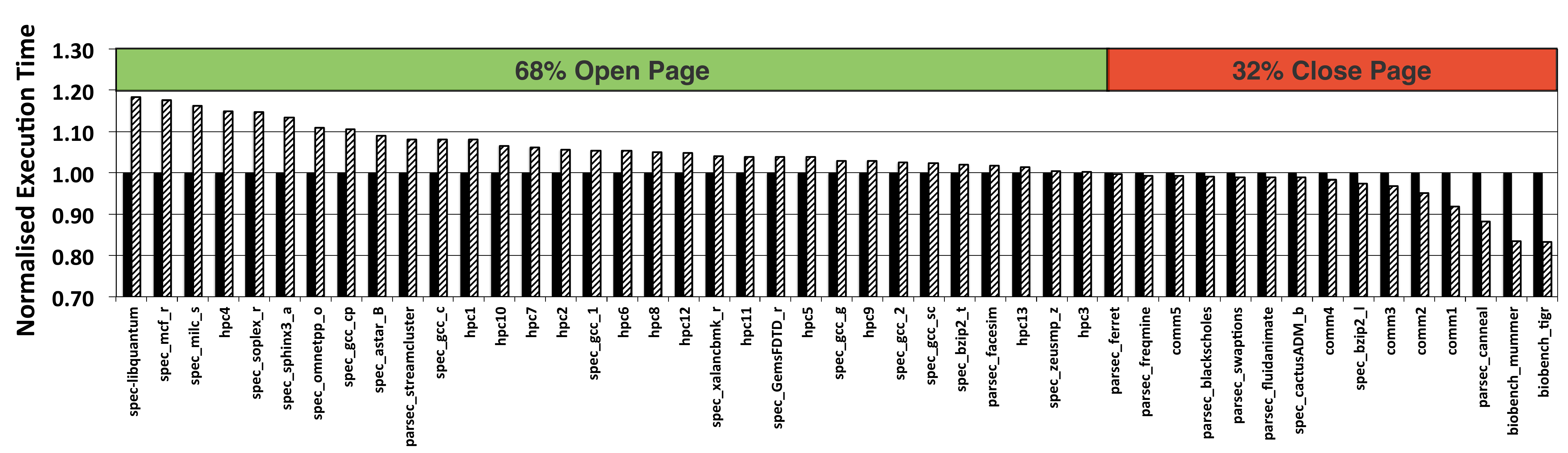}
\caption{Performance of Static Page Policies for standard workloads.}
\label{fig:motivation}
\end{figure*}

\begin{table}[!htb]
\centering
  \begin{tabular}{| c | c | c |}
	\hline
	\textbf{Page Policy} & \textbf{Page Hit} & \textbf{Page Miss} \\
	\hline
	\hline
	Open-Page & t\textsubscript{CL} & t\textsubscript{RCD} + t\textsubscript{CL} + t\textsubscript{RP}  \\
	\hline	
	Close-Page & t\textsubscript{RCD} + t\textsubscript{CL} & t\textsubscript{RCD} + t\textsubscript{CL}  \\
	\hline	
	Static Profiling & t\textsubscript{CL} & t\textsubscript{RCD} + t\textsubscript{CL}  \\
	\hline	
  \end{tabular}
  \caption{Cost of Page-hits and Page-misses when using different page closure policies.}
  \label{table:Cost_of_different_page_closure_policies}
\end{table}

{\bf Motivation:} Figure \ref{fig:motivation} depicts the normalised execution time of all the workloads (to open-page policy) that is used in this paper using open-page and close-page policy. The results show that around 68\% of workloads prefer the open-page policy while 32\% of workloads deliver a better performance using the close-page policy. According to this figure, a memory system that employs the open-page policy can save up to 18\%, in comparison with the close-page policy, when running `libquantunm' from SPEC benchmark and, at the same time, might lose up to 18\% when running `tigr' from BIOBENCH benchmark. Therefore, there is almost a 40\% of performance variation in the system depending on the static page policy that a memory controller employs. This motivates to start thinking about developing dynamic page policies that switch between open and close-page policy at run time based on the application access behavior. The \textit{Static Profiling} presented in Table~\ref{table:Cost_of_different_page_closure_policies} shows the cost of page-hits and page-misses when the memory controller selects \textit{the best} static page closure policy scheme for each workloads by static profiling of memory accesses. The static profiling provides a baseline to evaluate the performance of dynamic page closure policies discussed in this paper.

%improvement that a DRAM could achieve by the correct prediction of the page closure policy. Our experimental results over all the workload mixes will show that there is around 30\% potential performance improvements, over static open- and close-page policies respectively.

Motivated by this kind of observations, hybrid-page closure policies emerged. This type of page policies uses various prediction algorithms to switch dynamically between open- and close-page according to the application access behavior and improve performance.  Prediction accuracy and scalability with increasing memory size are the two main constraints when designing such page closure predictors. For most techniques in the literature, there is a linear relationship between the DRAM memory size and the required resources to monitor the memory access pattern. Thus, as the memory size grows, the required cost of page closure policy predictor grows and, as a result, the on-chip memory controller complexity and area increase. %On the other hand, in modern computer architectures, the memory controller is integrated on the die as a part of the processor. Thus, increasing the size of memory controller increases the die area of the processor which is not economic.

Overall, considering the scalability of emerging memory systems like HMCs, increasing the interest of using a large amount of DRAM instead of disk storage in servers and database analytic applications such as using 64~TB of DRAMs in RAMCloud \cite{website:RAMCloud,ousterhout2010case, ongaro2011fast} and using 150~TB of DRAMs by Facebook in memcache \cite{ongaro2011fast} all demand a scalable approach for efficient design. Section \ref{sec:HAPPY} introduces HAPPY as a compact encoding scheme to address the scalability problem of the page closure policy prediction technique for DRAM memory systems.

\section{HAPPY:Hybrid Address-based Page PolicY}
\label{sec:HAPPY}

\subsection{HAPPY -- Basic Principles}

HAPPY is a compact memory address-based encoding built on the observation that there is a strong correlation between physical address bits and the internal structure of DRAMs. Understanding the basic operation of DRAMs shows that one of the first steps to access to the DRAM structure is the address mapping process. During this process, the physical address bits provided by a core are translated to the corresponding channel, rank, bank, row and column of a DRAM device using a fixed and pre-defined address-mapping algorithm. Having a fixed translation mapping creates a strong correlation between physical address bits and the DRAMs structure. It means that, if some useful information can be extracted from the physical address bits after translation, it is possible to extract the same information before this stage.

All the page closure algorithms proposed so far focus on monitoring the memory access behavior after the translation phase. In general, these techniques use different performance counters in a channel, bank or row basis to monitor page hits/conflicts, the time that a row could be kept open etc. HAPPY proposes a novel binary-encoding scheme with performance counters storing page closure history directly from the physical address bits. In the other words, HAPPY introduces one predictor per physical address bit to forecast the page closure policy of each row in the memory system according to the run-time memory accesses.

Sections \ref{sec:happy_access} and \ref{sec:happy_time} show how HAPPY can be applied to the two most main page closure categories: access-based and time-based techniques. We illustrate one traditional and one state-of-the-art technique to demonstrate how the HAPPY encoding can be applied to different systems with different implementation characteristics. The methodology proposed in this paper can be applied to other aspects of a DRAM structure as well.   

\subsection{HAPPY -- Access-based Prediction}
\label{sec:happy_access}

\begin{figure}[!htb]
\centering
\includegraphics[scale=0.3]{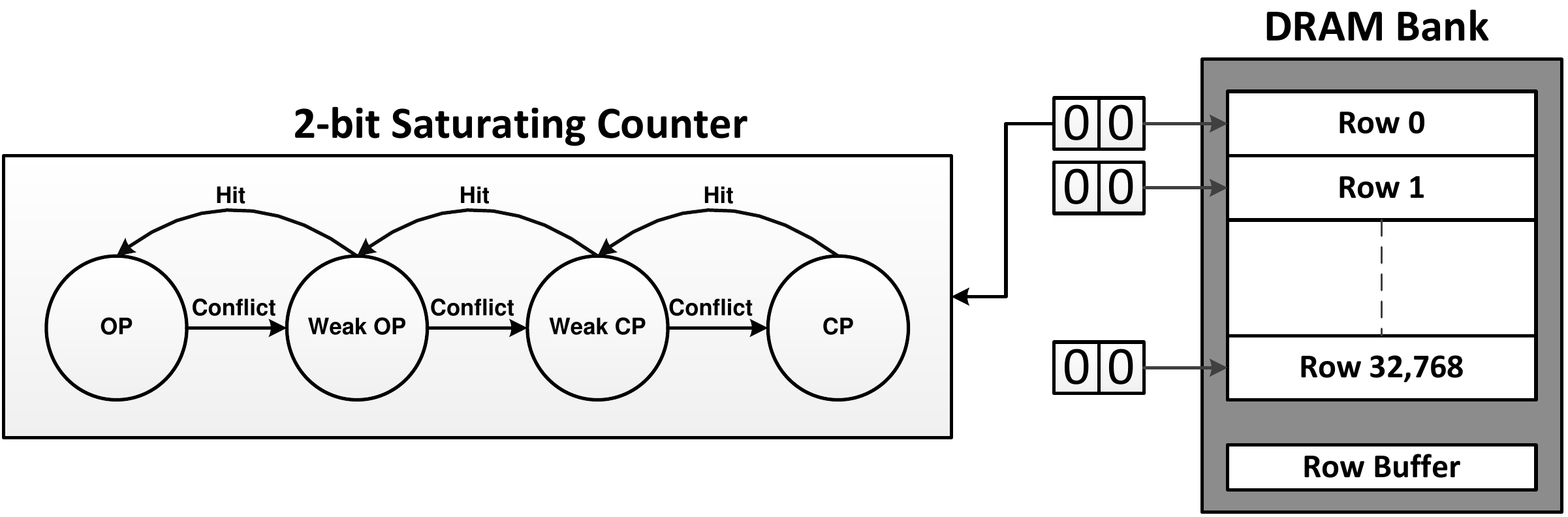}
\caption{Basic structure of Hybrid Page Policy.}
\label{fig:Hybrid_basic_Structure}
\end{figure}

To demonstrate how HAPPY can be applied to access-based algorithms we select the traditional hybrid-page policy algorithm. This employs simple, saturating counters to monitor the memory access pattern behavior and dynamically switch between open- and close-page policy at run time. Figure~\ref{fig:Hybrid_basic_Structure} depicts the basic structure of such page closure policy predictors.

In this technique, one saturating counter (e.g.\ a 2-bit counter initialized to zero - open-page policy) is assigned to each row of a DRAM bank. Every time that a row-miss happens the corresponding counter is incremented. Whenever a row-hit occurs the counter is decremented. For each memory request, the accessed row's counter value determines the page closure policy; if the value is 0 or 1 the open-page policy is predicted and if the value of the counter is 2 or 3 the close-page policy is predicted. 
%Our experimental results will show that this simple technique delivers a better performance than static page policies almost for most workloads. 
However, having a counter for each row in a DRAM device imposes a high area and power overhead to the memory system. For example, a 4~GB DRAM memory system with 1 channel, 2 ranks, 8 banks and 32,768 rows per bank, require 524,288 counters, which is not scalable (analysis presented in Section \ref{sec:Results_and_Discussions}).

\begin{figure}[!htb]
\centering
\includegraphics[scale=0.32]{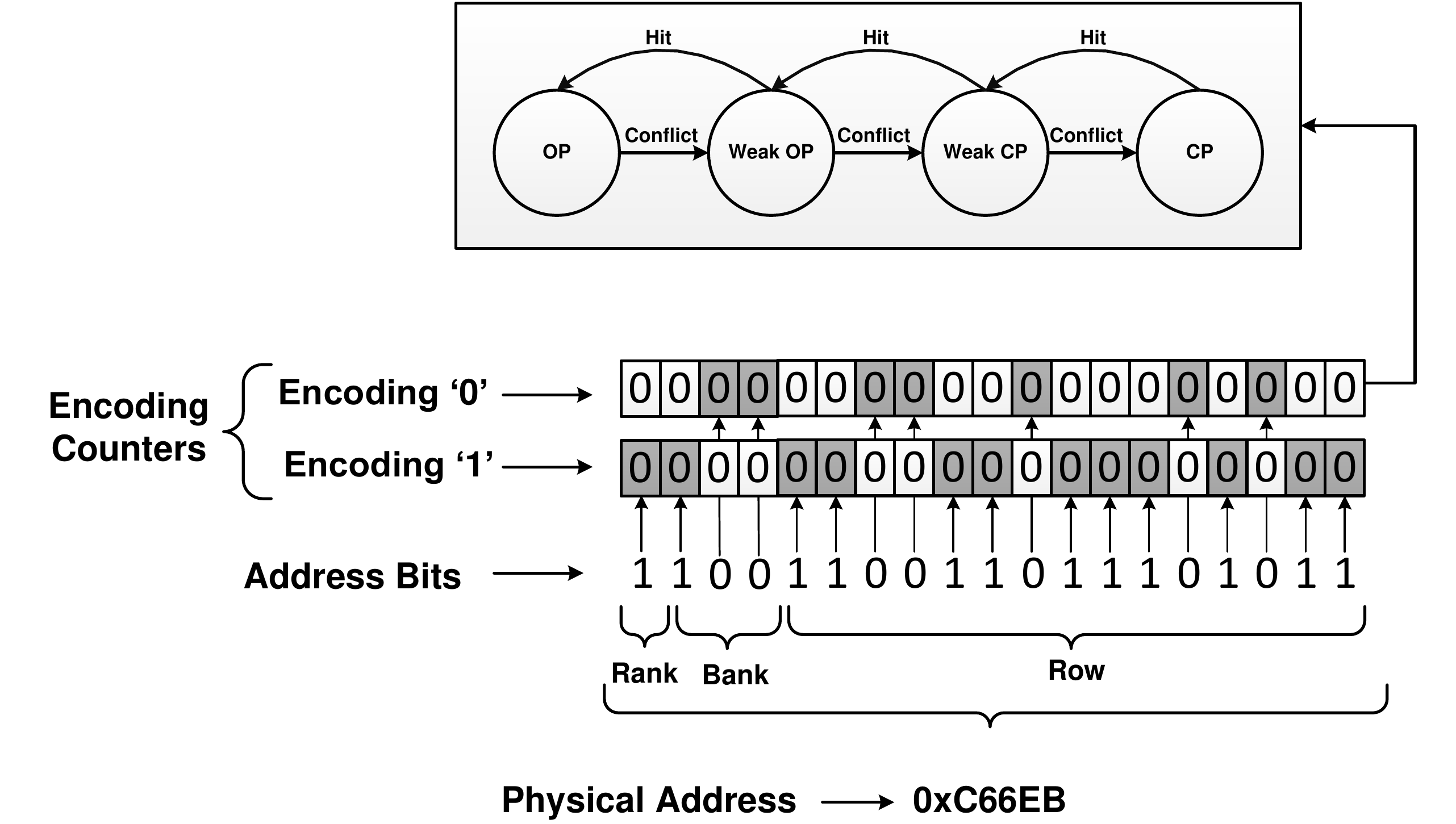}
\caption{HAPPY implementation of Hybrid Page Policy.}
\label{fig:HAPPY_Hybrid_basic_Structure}
\end{figure}

Figure~\ref{fig:HAPPY_Hybrid_basic_Structure} depicts the HAPPY implementation of a Hybrid page policy. The binary representation of the requested physical address is a pattern of zeros and ones. HAPPY dedicates two encoding counters per physical {\em address bit} location: one counter to monitor the position when its value is ‘1’ and one to monitor it when it is ‘0’. Training of these counters is similar to the original implementation of Hybrid; that means for every page conflict the counter corresponding to each physical address bit is incremented and for every page hit the same counter will be decremented. Considering the page-hits and page-misses happen in the row basis then there is no need to monitor all the available physical address bits. Thus, the corresponding physical address bit to columns and cache-lines offset will not be used. This reduces even further the implementation costs of HAPPY.

Having done this, each physical address bit correlates with the possibility of getting page hit/conflict depending on the value of that bit. Therefore, for a given physical address the possibility of getting page hit/conflicts can be calculated simply by considering at all the participant bit’s counter values in the requested address and using one of the following techniques: Majority vote or Aggregation. 

\begin{figure}[!htb]
\centering
\includegraphics[scale=0.33]{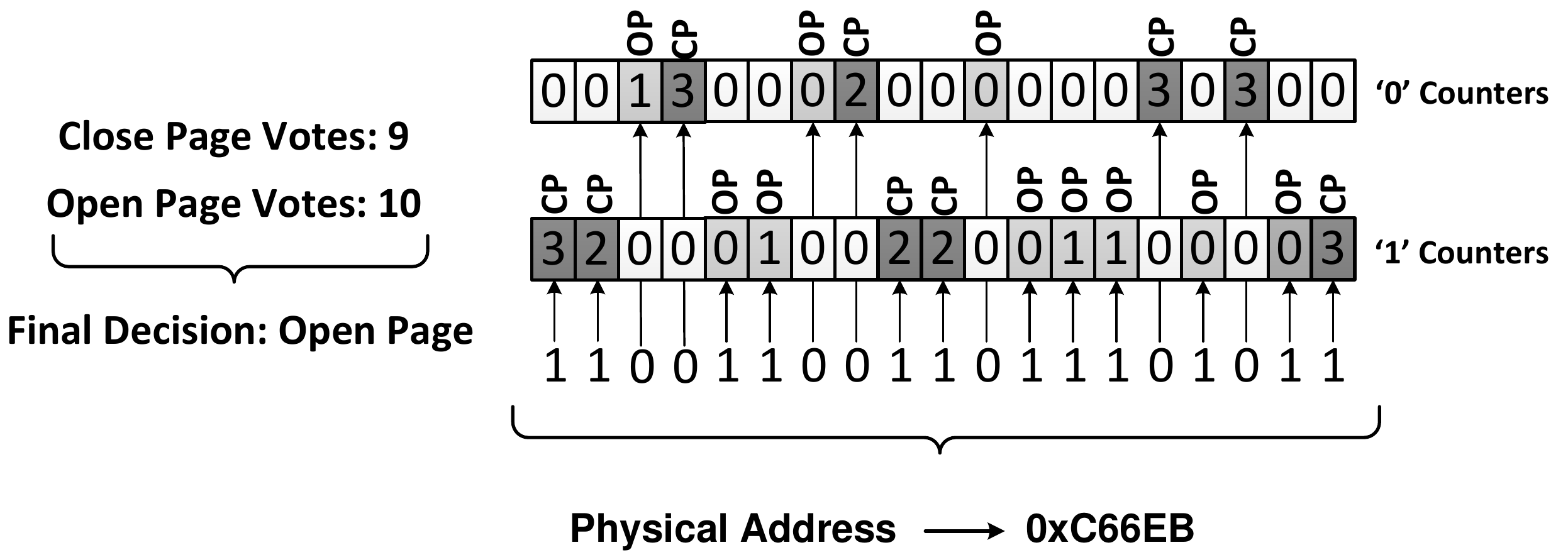}
\caption{Example of HAPPY Majority vote decision.}
\label{fig:HAPPY_Majority_Vote}
\end{figure}

{\bf Majority vote:} Figure~\ref{fig:HAPPY_Majority_Vote} explains this scheme using a simple example. Each physical address bit has a counter which has its own standalone vote to choose an open- or close-page policy for the requested physical address. The page closure policy vote of each bit can be extracted by looking at the more significant bit of the saturating counters. If this bit is ‘0’ there is an open-page policy vote and if the value is ‘1’ there is a close-page policy vote. As the name of decision implies, the final vote is determined by the majority vote across all the counters.

{\bf Aggregation:} the final page closure policy decision can also be determined by comparing the aggregation of the all the counters values, Equation~\eqref{eq:AggregationDecision}, against a certain threshold, Equation~\eqref{eq:Aggregationthreshold}.

\begin{equation}
\begin{split}
&if \sum AddressBitCounters < Threshold \to Open \ Page \\
&else \quad \quad \quad \quad \quad \quad \quad \quad \quad \quad \quad \quad \quad \quad \  \ \to Close \ Page
\end{split}
\label{eq:AggregationDecision}
\end{equation}

\begin{equation}
Threshold = \frac{AddrBitsWidth\times CounterValue}{2}
\label{eq:Aggregationthreshold}
\end{equation}

The experimental results show  similar results using either majority or aggregation. Thus only the majority vote decision scheme is used in the final experimental results.

\subsection{HAPPY -- Time-based Prediction}
\label{sec:happy_time}

\begin{figure}[!htb]
\centering
\includegraphics[scale=0.39]{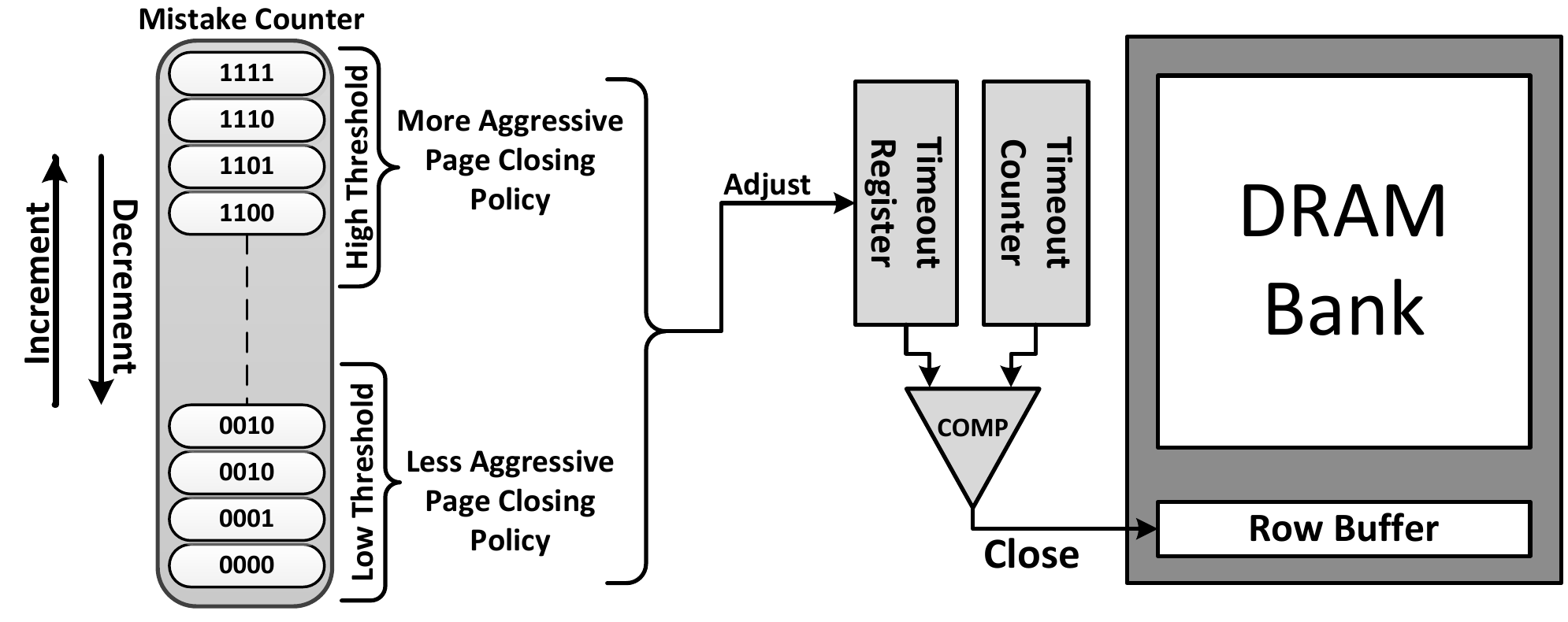}
\caption{Intel-adaptive page policy predictor basic structure.}
\label{fig:Intel-adaptive basic structure}
\end{figure}

As a case study to show how HAPPY can be applied to a time-based prediction algorithm we chose the Intel-adaptive open-page policy \cite{dodd2006adaptive,website:inteladaptive} employed by the Intel Xeon X5650 \cite{Intelweb}. The basic structure of such a page closure policy is presented in Figure~\ref{fig:Intel-adaptive basic structure}.

The integrated memory controller used in this Intel processor can be configured at boot time to employ one of the following three different page closure policy schemes: close-page, fixed-open and Intel-adaptive open page. The \textit{fixed-open page policy} keeps a row open for a fixed period of time and closes it after that. The Intel-adaptive scheme is an advanced version of the fixed-open schemes. Similar to the fixed-open policy, in this structure, each row buffer within a bank has a Timeout Counter (TC) and a Timeout Register (TR). A row will be kept open until TC reaches the TR and then closed. However, the initial TR might not be a suitable value for all the benchmarks. Thus, the Intel-adaptive scheme provides a technique to update the TR at run time using a 4-bit Mistake Counter (MC). Whenever a page conflict happens that could have been a page-empty, since there was enough time to precharge the last accessed row, then the MC is decremented. Whenever a page empty could have been a page-hit, since the row being accessed is the same as the last accessed row in that bank, then the MC is incremented. After a specific time interval the MC will be checked against a pre-predefined low and high threshold to see if either a less or more aggressive page closure policy is required. If the MC is higher than the high-threshold then the TR will be incremented to keep the accessed row open for a longer period and if the MC is lower than the low-threshold the TR will be decremented to close the accessed row sooner. 

\iffalse
Our experimental results shows that Intel-adaptive open page policy delivers the best execution time across all the investigated prediction algorithms in this paper. However, the current implementation of this scheme only considers one timeout register for all the rows within a bank. Ideally, we would like to keep track of desired timeout value for each row in a DRAM device. However this would be extremely costly to implement in a memory controller (e.g. about 1~MB extra storage for a 4~GB DRAM).
\fi

\begin{figure}[h!]
\centering
\includegraphics[scale=0.34]{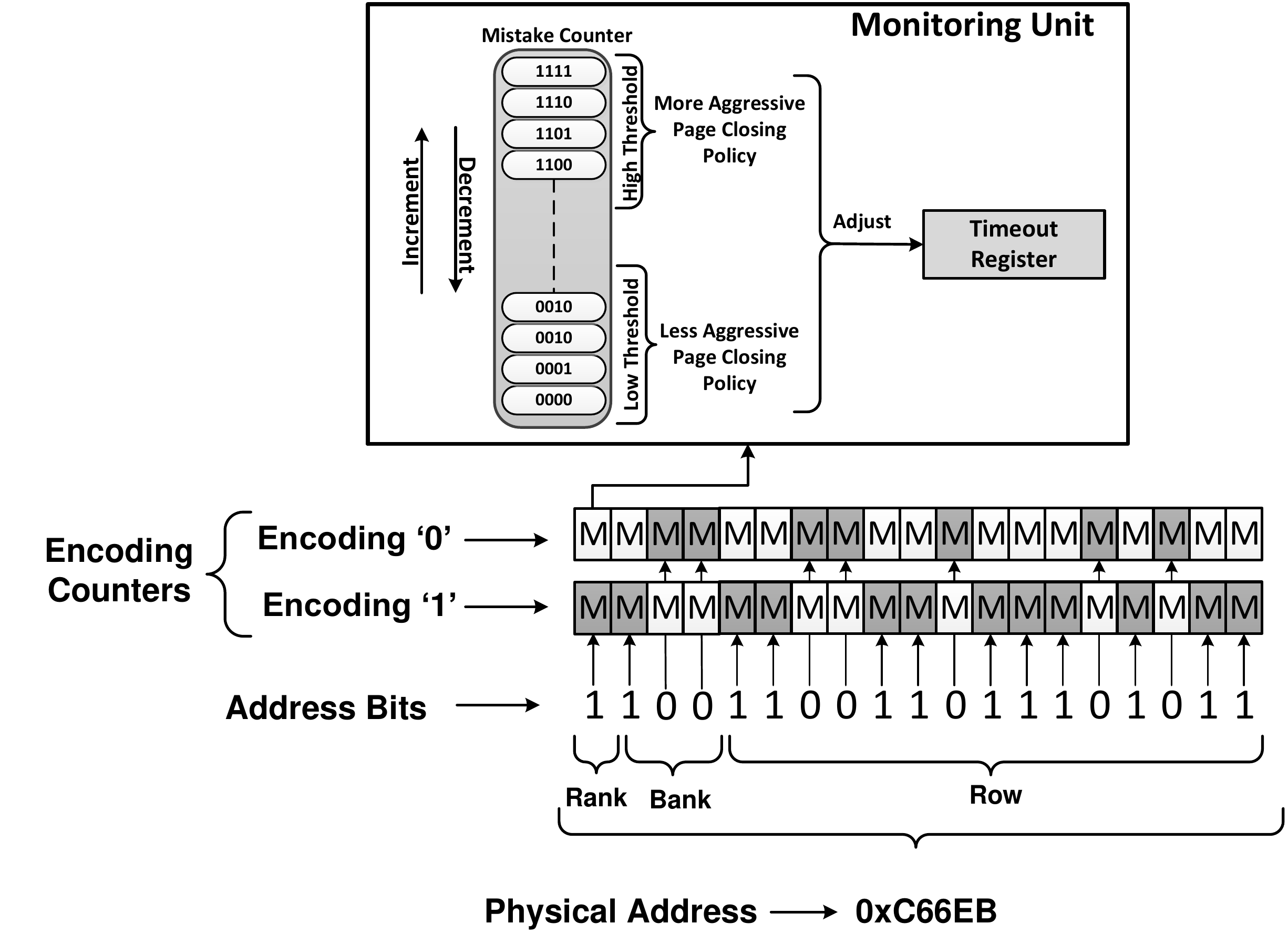}
\caption{HAPPY implementation of Intel-adaptive page policy predictor.}
\label{fig:HAPPY_Intel_Adaptive structure}
\end{figure}

Figure~\ref{fig:HAPPY_Intel_Adaptive structure} depicts the HAPPY implementation of the Intel-adaptive open page policy. This time the aim is to extract the timeout value for each row to be kept open from the physical address bits.  We use the same methodology explained in the previous section; the only difference is that instead of using simple saturating counters a Monitor Unit is dedicated to each physical address bit location. Each monitoring unit includes a MC and a TR with the same function as the original implementation of Intel-adaptive page policy. A global timeout counter is still required to keep track of row closing and opening times on a per bank basis.

Updating the MCs is as before but this time it is applied per physical address bit basis rather than per bank. Moreover, instead of having a global timeout register per bank, the time period that a row can be kept open will be calculated from the aggregation of all the participant bits for an accessed physical address.

To make a fair comparison between HAPPY and original implementation of this page policy, the size of the MC is chosen as before (e.g.\ 4-bit) and the size of TR in each physical bit location is chosen small enough that the maximum time that a row can be kept open is equal to having one global TR per bank.

\subsection{HAPPY -- Intuition}

\begin{itemize} 

\item {\bf Observation:} The main intuition behind HAPPY is based on the observation that addresses that are spatially close together tend to have a similar page-closure policy preference. HAPPY is devised to exploit such behavior by fine grain monitoring of physical address bits behavior. 

\item {\bf Ensemble Methods:} Although HAPPY was developed by analysis and observation of the experiments, we believe that there are Machine Learning principles that can justify the intuition behind HAPPY.  A mathematical/theoretical framework that can explain HAPPY is that of Ensemble Methods \cite{brown2010}. The family of algorithms categorized as Ensemble methods combines multiple (normally simple) predictors. The theory explains how combining such predictors, it can be obtained a much improved predictor provided certain diversity properties among the predictors are met. Random forest, and neural networks are examples of very successful prediction algorithms part of the ensemble family. This paper addresses an online learning scenario and uses a fixed number of predictors with a non-linear combination function. When applying our technique to Intel-adaptive, we generate a pair of simple predictors per physical address bit and solve a regression problem. Each pair of predictors is trained using a different single physical bit (different features) and each member of the pair is trained using only Zero or One occurrences (different dataset); mechanisms that can improve diversity. In Happy, we use two counters per physical address bits; e.g. 4GB is represented with 38 counters. If we were to limit ourselves to only a binary decision, then the maximum number of possible decisions that can be stored would be 2 to the power of 38. If we increase to 8GB, then we use 40 counters, and thus the maximum number of possible decisions that can be stored would be 2 to the power of 40. Thus as we increase the size of memory we are also increasing the number of possible decisions representable.

\end{itemize}

%\subsection{HAPPY – Why it works?}
%To do!

%\subsection{HAPPY – Further Possible improvements}
%This paper is mainly intended to explain the basic principles behind HAPPY and how it can significantly improve the scalability of the prediction algorithms. However, a significant reduction in the cost of implementation provides further opportunity to design a more efficient and accurate predictor. 

%For instance, our experimental results show that increasing the size of saturating counters from 2-bit to 4-bit for the first case study can improve the prediction accuracy and, as a result, the performance of memory system. However, for a 4~GB DRAM that requires around 500K saturating counters; two extra bits per counter impose a significant area and power overhead to the system which is impractical. Now, considering that HAPPY can reduce the cost of implementation by 13,000$\times$ (i.e. for a 4~GB DRAM) without harming the performance there is enough space to increase the counter size with negligible overhead.  

\section{Evaluation Methodology}
\label{sec:evaluation}

Page closure prediction algorithms for DRAMs are sensitive to the application memory access patterns. To address this, we carry out an extensive evaluation, described as follows:

{\bf Simulator:} USIMM \cite{chatterjee2012usimm}, a detailed memory system simulator, is used as our main simulation platform. We extended USIMM to support five different, existing page closure policies (i.e.\ Open-Page, Close-Page, Hybrid, Fixed-Open and Intel-adaptive open page) plus the two implementations of HAPPY which are described in Section~\ref{sec:HAPPY} (i.e.\ Hybrid-HAPPY and Intel-adaptive-HAPPY). The scheduling algorithm in the memory controller is FR-FCFS; first ready, first come first serve. We evaluated HAPPY based on different memory configurations, 2 GB for single-thread and 4 GB for multithread workloads. To increase the memory congestion we configured USIMM with 1 channel and 1 rank. The baseline USIMM system configuration parameters are captured in Table~\ref{table:usimm_config_parameter}. 

% as well as the official workloads part of the Memory Scheduling Championship (MSC) \cite{chatterjee2012msc}. 
%We evaluate all the prediction algorithms on one- and four-channel memory systems to investigate their sensitivity to the memory configuration. 

\begin{table}[!htb]
\centering
  \begin{tabular}{| >{\centering\arraybackslash}m{4cm} | >{\centering\arraybackslash}m{3.5cm} |}
	\hline
	\multicolumn{2}{|c|}{\textbf{Processor}} \\	
	\hline
	Clock Speed & 3.2GHz \\
	\hline
	 Pipeline depth & 10\\
	 \hline
	 ROB size & 32\\
	 \hline	 
	 
	 %\hline	
	%\rowcolor[gray]{.8} \multicolumn{2}{|c|}{\textbf{Cache Hierarchy Configuration}} \\	
	%\hline
	%L1 I-cache & 32K/4-way, Private\\
	%\hline
	%L1 D-cache & 32K/4-way, Private\\
	%\hline
	%L2 Cache & 2MB/16-way, Shared\\
       	%\hline	
	
	\hline
	\multicolumn{2}{|c|}{\textbf{DRAM Parameters}} \\	
	\hline
	DRAM Size & 2 - 4 GB\\
	\hline
	 Bus Speed & 800MHz\\
	 \hline
 	 Configuration & 1Channel,1Rank,8Banks\\
	 \hline
	% Ranks per channel & 1\\
	 %\hline
	 %Bank per rank & 8\\
	 %\hline
	 Row per bank & 65,536\\
	 \hline
	 Cache line per row & 128\\
	 %\hline
	 %Cache line size & 64 Byte\\
	\hline
  \end{tabular}
  \caption{Simulation paramteres.}
  \label{table:usimm_config_parameter}
\end{table}

\begin{figure}[h!]
	\centering
	%\begin{subfigure}
	\subfloat[Mapping 1: Maximise row-buffer locality]{\includegraphics[scale=0.3]{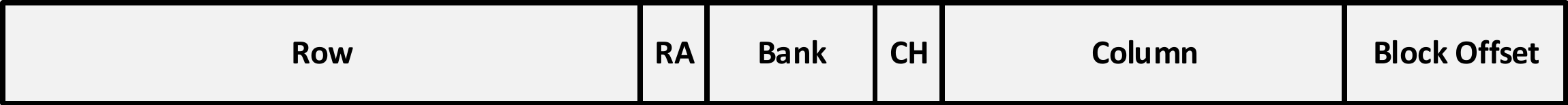}\label{fig:mapping_1}} \\
	%\subfloat[Mapping 2: Maximise memory access parallelism]{\includegraphics[scale=0.3]{Figures/Address_Mapping_2.pdf}\label{fig:mapping_2}} \\
	\subfloat[Mapping 3: Permutation-based Page Interleaving \cite{zhang2000permutation}]{\includegraphics[scale=0.3]{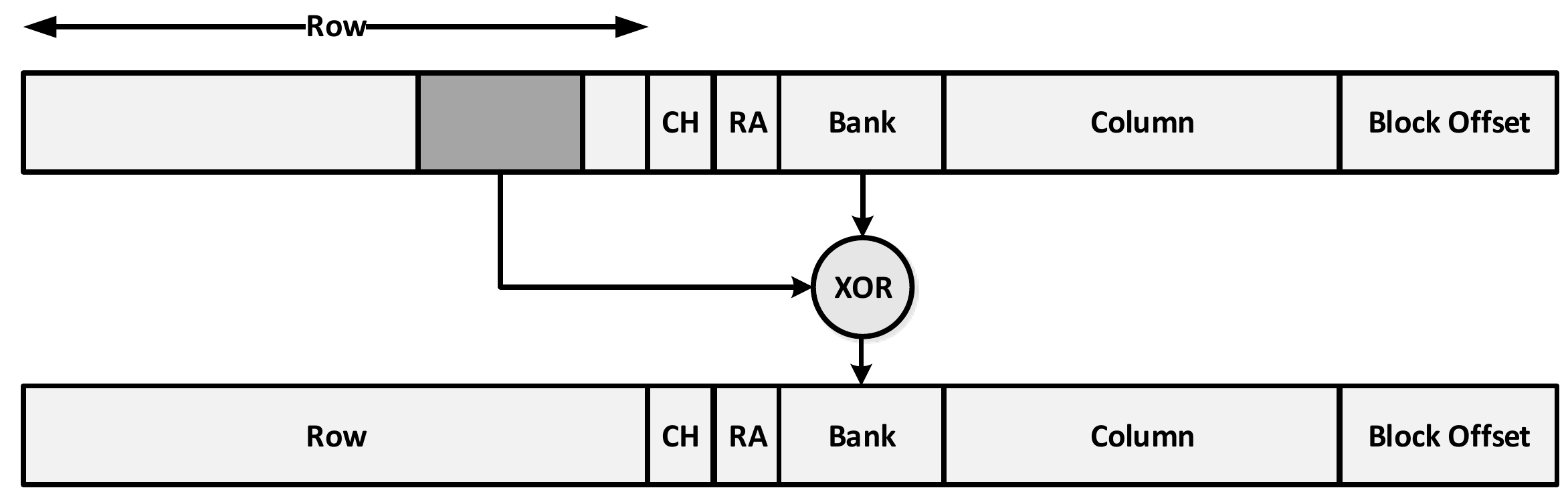}\label{fig:mapping_3}} \\
	\subfloat[Mapping 4:Minimalist Open-Page Scheme \cite{kaseridis2011minimalist}]{\includegraphics[scale=0.3]{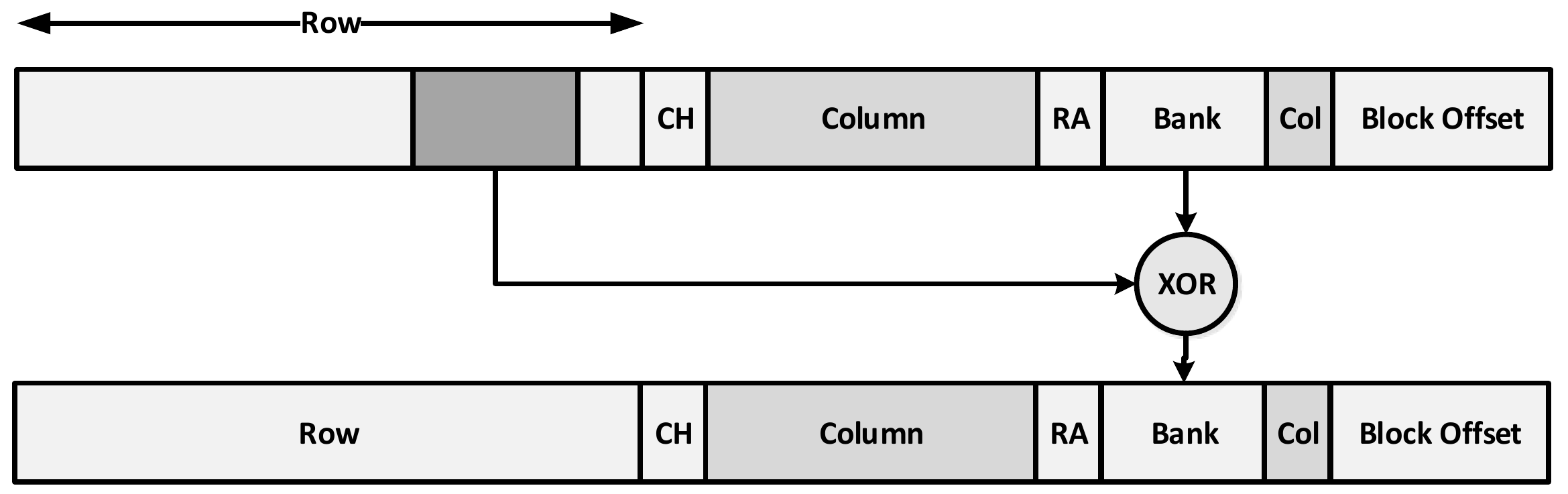}\label{fig:mapping_4}} 		

	\caption{Different Address Interleaving schemes.}	
	\label{fig:Address_Mapping_fig}
	%\end{subfigure}
\end{figure}

{\bf Address Mapping Schemes:}  The number of page conflicts in DRAMs and as a result the memory performance is susceptible to the memory address mapping scheme. The experiments consider 3 different address mappings presented in Figure~\ref{fig:Address_Mapping_fig}. The first mapping 1 is a standard mappings to maximize row buffer locality. The next two address interleaving policies are state-of-the-art schemes proposed by Kaseridis~\textit{et al}.\ \cite{kaseridis2011minimalist} and Zhang~\textit{et al}.\ \cite{zhang2000permutation}. The proposed mapping by Zhang~\textit{et al}.\ \cite{zhang2000permutation} XORs part of the row address bit with the banks address bits to produce a new bank index (see Figure~\ref{fig:mapping_3}). Kaseridis~\textit{et al}.\ \cite{kaseridis2011minimalist} extend this technique by producing the column index using different section of physical address bits (Figure~\ref{fig:mapping_4}). Both techniques aim to reduce page conflicts in DRAMS. Our experiments shows that the minimalist open page policy (Mapping 3) performs better for most of the workloads. Therefore, this address mapping scheme is employed in all the experiments. Focusing on the best page closure policy (i.e.\ Intel-Adaptive-HAPPY), we also report the sensitivity of HAPPY for all the three address mappings schemes.

{\bf Workloads:} The workloads include a wide range of memory intensive applications (i.e.\ 48 workloads) from different  benchmark suites (PARSEC \cite{bienia2008parsec}, SPEC \cite{dixit1991spec}, BIOBENCH \cite{albayraktaroglu2005biobench}, HPC and COMMERCIAL) and representative regions of interest for each application. Table~\ref{table:workloads_characterisation} lists the workloads and their corresponding benchmark suites.

\begin{table*}[!htb]
\centering
  \begin{tabular}{| >{\centering\arraybackslash}m{3cm} | >{\centering\arraybackslash}m{3.5cm} |  >{\centering\arraybackslash}m{3.5cm} | >{\centering\arraybackslash}m{3.5cm} |}
	\hline
	\multicolumn{4}{|c|}{\textbf{Benchmark Suites}} \\	
	\hline
	\multicolumn{2}{|c|}{\textbf{SPEC}} & \textbf{PARSEC} & \textbf{COMMERCIAL} \\	
	\hline
	(a) GemsFDTD\_r & (k) astar\_B & (u) canneal & (D1) comm1\\	
	\hline
	(b) bzip2\_l & (l) bzip2\_t & (v) streamcluster & (D2) comm2\\	
	\hline
	(c) cactusADM\_b & (m) gcc\_1 & (w) blackschols & (D3) comm3\\	
	\hline
	(d) gcc\_2 & (n) gcc\_c & (x) facesim & (D4) comm4\\	
	\hline
	(e) gcc\_cp & (o) gcc\_g & (y) ferret & (D5) comm5\\	
	\hline
	(f) gcc\_sc & (p) mcf\_r & (z) fluidanimate & \textbf{BIOBENCH}\\	
	\hline
	(g) milc\_s & (q) omnetpp\_o & (A) freqmine & (E) mummer\\	
	\hline
	(h) soplex\_r & (r) sphinx3\_a & (B) swaption &  (F) tigr\\
	\hline
	(i) xalancbmk\_r & (s) zeusmp\_z &  \textbf{HPC} & \cellcolor[gray]{0.2} \\
	\hline
	(j) libquantum & (t) leslie & (C) hpc1 - hpc13 & \cellcolor[gray]{0.2} \\
	\hline
  \end{tabular}
  \caption{Standard workloads and benchmark suites. }
  \label{table:workloads_characterisation}
\end{table*}

The USIMM simulator can run arbitrary multi-application workloads using multiple traces. To increase the variety of memory access patterns, we set up USIMM for multi-applications to produce 22 random workload mixes; a combination of 4-thread, 8-thread and 16-thread applications. Table \ref{table:multicore_workloads} listed these multi-core workloads considering the prefix of single thread workloads presented in Table \ref{table:workloads_characterisation}. Overall the experiments consider 70 workload mixes.

{\bf Memory Footprint (MF):} To evaluate the performance of page closure predictors a careful study has to be carried out otherwise the performance and accuracy numbers might be misleading. For instance, if an application targets a very small portion of memory then it might be possible to predict its behavior using very small number of performance counter whereas if the application accesses all over of the memory space then it might be more difficult to keep track of the application access pattern with only a few counters (e.g. HAPPY). To have a fair evaluation methodology we made sure that the memory traces cover a wide range of access pattern. To this aim, we monitored the total physical pages accessed (Memory Footprint) per application at run time and we can confirm that our single thread applications have the average MF of 30\% (up to 97\%), our 4-thread workloads have the average MF of 50\% (up to 75\%), our 8-thread workloads have the average MF of 70\% (up to 85\%) and our 16-thread workloads have the average MF of 95\% (up to 99.8\%).

\section{Results and Discussions}
\label{sec:Results_and_Discussions}

This section analyzes the different page closure policy prediction schemes compared with using HAPPY by looking at execution time, accuracy and scalability. Before jumping to the result graphs the following summary might be helpful: \\

\begin{itemize}

\item The HAPPY implementation of Hybrid page policy is called
  Hybrid-HAPPY for brevity.

\item The HAPPY implementation of Intel-adaptive open-page policy is
  called Intel-adaptive-HAPPY for brevity.

%\item Figure \ref{fig:Accuracy_graph} presents the prediction accuracy in terms of page hit and misses.  

\item The results in Figure~\ref{fig:Final_Results} and Figure~\ref{fig:hpc_results}-\ref{fig:multithread_thread_results} are
  normalized to the `static profiling'; the lower the bar
  the better performance.

%\item Overall, the HAPPY implementations perform similar to (or better than) state-of-the-art policies while reducing significantly the hardware overheads for dynamic page closure policies. Note that while average geometric mean performance improvements are small (in average 5\% and 8\% over open-page and close-page respectively), HAPPY requires a minimum 5$\times$ less storage overhead than earlier techniques. Unlike prior proposals, the hardware overhead of HAPPY is scalable with the DRAM memory size.

%\item Figures \ref{fig:Mapping_Sens_Hit_graph} and \ref{fig:Mapping_Sens_Miss_graph} present the prediction accuracy of the best predictors (Intel-adaptive and Intel-Adaptive-HAPPY) analyzing the effect of different address mapping schemes. We observe that Intel-Adaptive-HAPPY using fewer hardware resources delivers a slightly better accuracy (i.e.\ 2\%) than Intel-Adaptive for all three address mapping schemes.

\end{itemize}

\subsection{Prediction Accuracy} 

Understanding the prediction accuracy for the different types of page closure predictors has its pitfalls. For instance, prediction accuracy in the case of Hybrid predictors is straight forward as the prediction outcome is either opening or closing a page (binary classification). However, in the case of the Intel-adaptive technique the accuracy needs to be described based on the timeout value (regression). Consider a scenario where a page has to be open for 40 clock cycles to get a page hit and Intel-adaptive predicts 39 clock cycle. In this case the prediction accuracy should be calculated differently. 

\begin{figure}[!htb]
\centering
\includegraphics[scale=0.2]{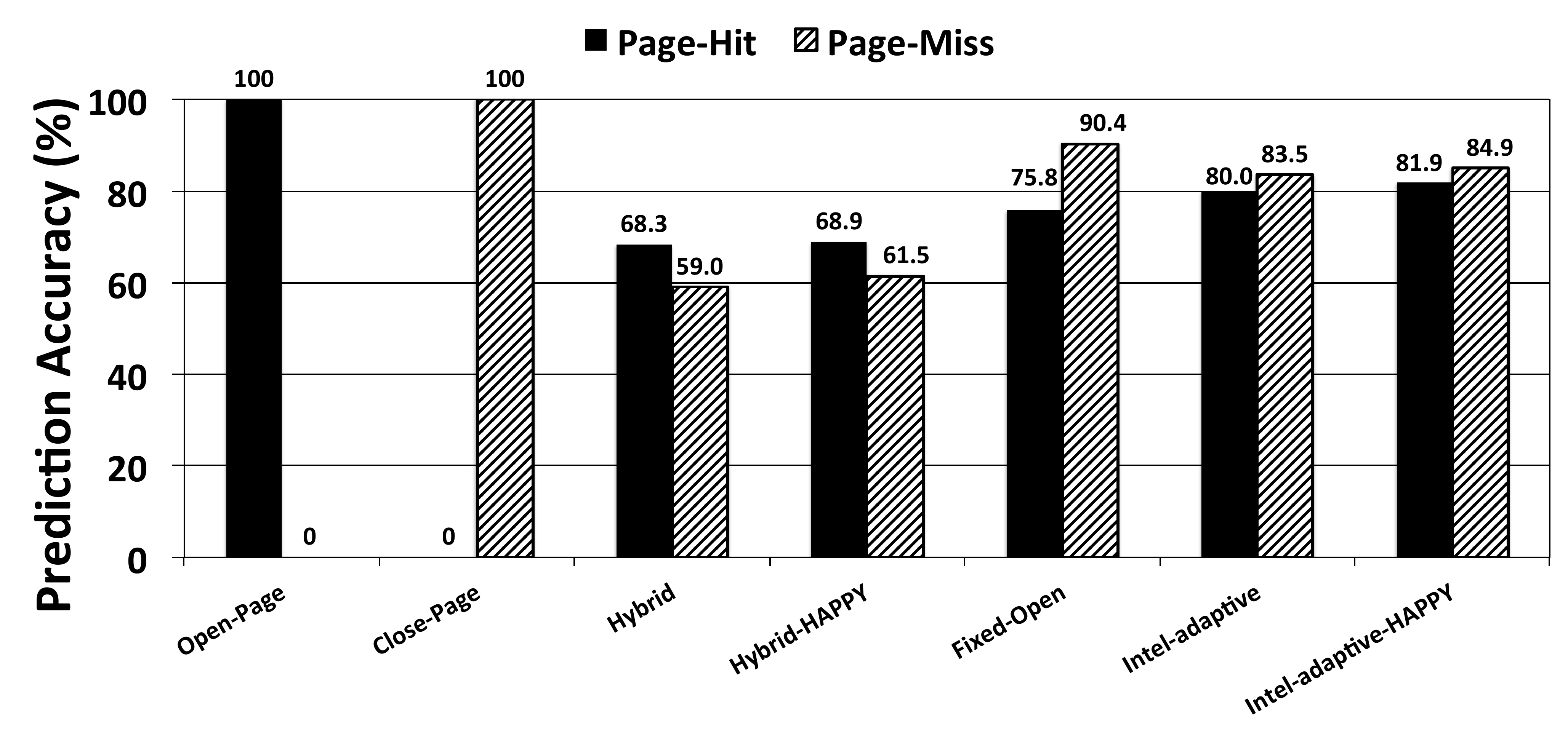}
\caption{Prediction accuracy for different predictors.}
\label{fig:Accuracy_graph}
\end{figure}

\begin{figure*}
\centering
\includegraphics[scale=0.15]{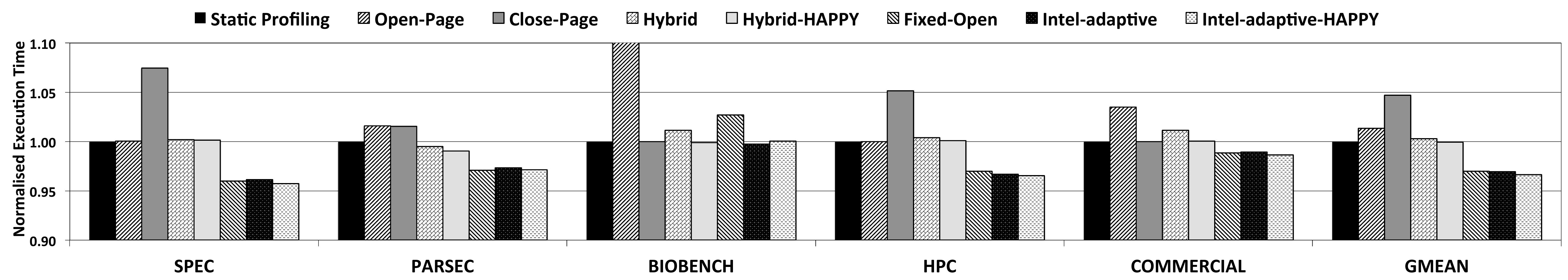}
\caption{Average relative performance to static profiling for all the single-thread workloads.}
\label{fig:Final_Results}

\vspace{5 mm}

\centering
\includegraphics[scale=0.15]{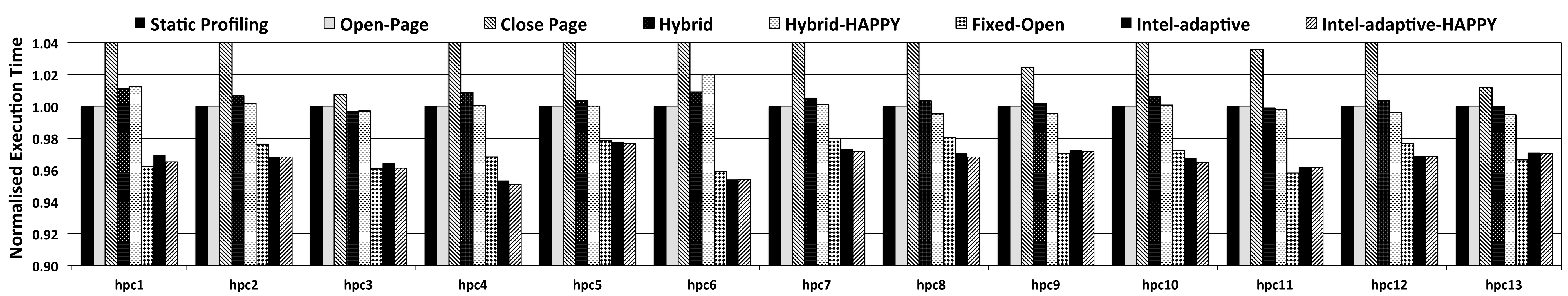}
\caption{Relative performance to static profiling for HPC workloads.}
\label{fig:hpc_results}

\vspace{5 mm}

\centering
\includegraphics[scale=0.15]{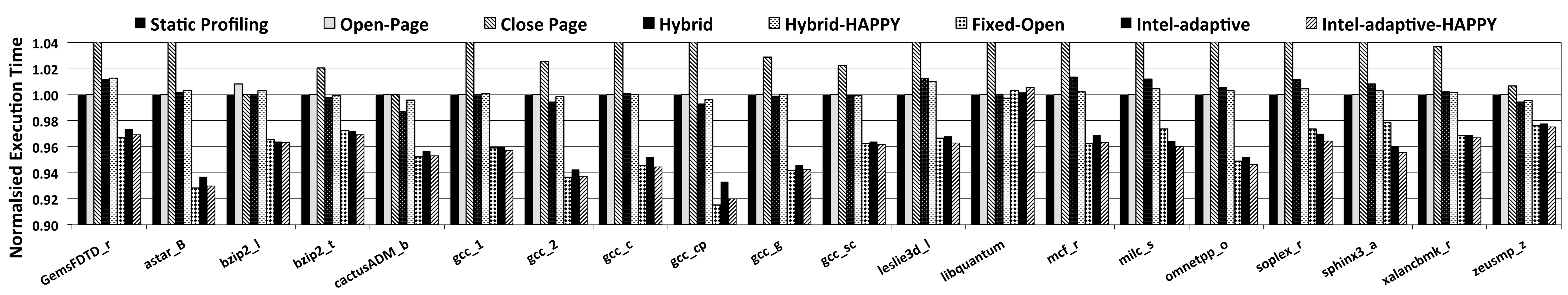}
\caption{Relative performance to static profiling for SPEC workloads.}
\label{fig:spec_results}

\vspace{5 mm}

\centering
\includegraphics[scale=0.15]{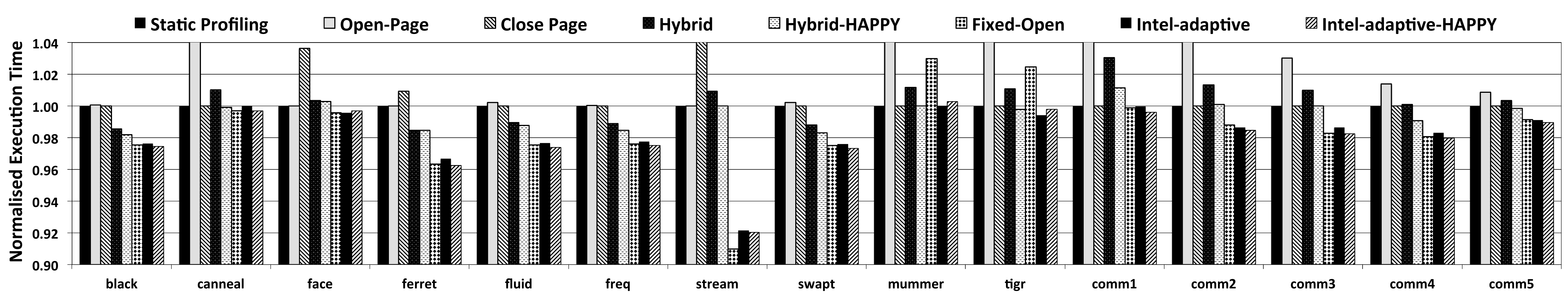}
\caption{Relative performance to static profiling for PARSEC, BIOBENCH and Commercial workloads.}
\label{fig:parsec_results}

\vspace{5 mm}

\centering
\includegraphics[scale=0.15]{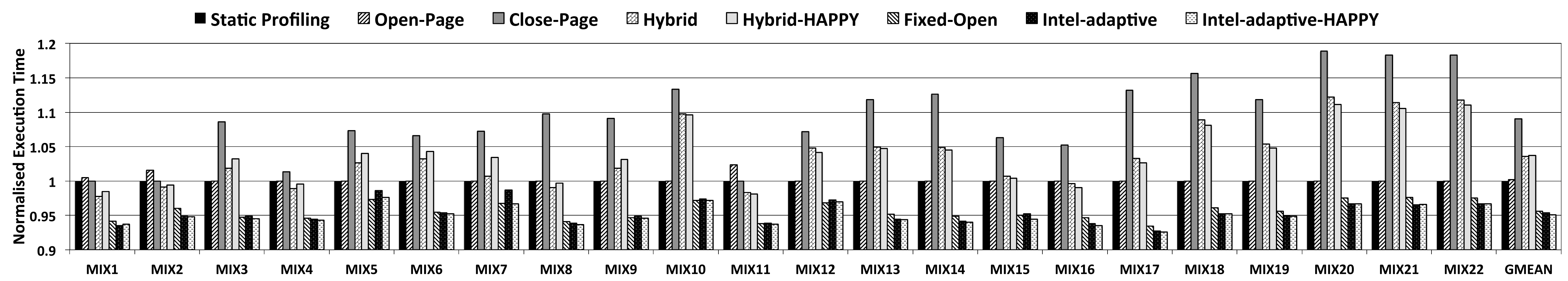}
\caption{Relative performance to static profiling for Multi-Core workloads.}
\label{fig:multithread_thread_results}
\end{figure*}

To have a fair evaluation across all the predictors with a different nature of prediction, we calculate the prediction accuracy based on the Page-Hit and the Page-Miss prediction outcome. In fact, the main purpose of using page policy predictors is to increase the page hits and reduce the page misses in the DRAM. Thus, we calculate the \textit{Oracle page-hits} (maximum possible page-hits when having a perfect predictors) and \textit{Oracle page-misses} (minimum possible page-misses when having a perfect predictor) and evaluate the actual page-hits and page-misses occurred during execution time of each workloads against the oracle numbers. Figure~\ref{fig:Accuracy_graph} presents the prediction accuracy (GMEAN) of different predictors across all the workloads evaluated in this work. The open-page and close-page policies deliver the maximum prediction accuracy for page-hits and page-misses, respectively. This happens because an open-page policy leaves all the page open and then it can cover all the possible page-hits in the system and non of the page misses. A close-page policy behaves similarly but in the opposite scenario.
The hybrid-page policy delivers a moderate page-miss and page-hit prediction accuracy (around 60\%). The Intel-adaptive and fixed-open both deliver a good prediction accuracy for both page-hits (80\% and 75.8\%) and page-misses (83.5\% and 90.4\%) respectively. Overall, the HAPPY implementation of both Intel-adaptive and hybrid are slightly more accurate than the original implementation. This prediction accuracy numbers justify the execution time presented in Figure~\ref{fig:Final_Results}. Also, from the accuracy results this can be concluded that the page-hit prediction accuracy has a higher impact on the overall execution time than the page-miss prediction accuracy. 

%Our results show that the Workload-Partitioning predictor is fairly sensitive to its initial parameters such as observation window and the current results show that it can deliver a good prediction of page-miss prediction (97.7\%) but it may not be accurate in the case of page-hits. 

\subsection{Performance Analysis}
Figure~\ref{fig:Final_Results} summarizes the performance of different prediction algorithms, normalized to the `Static Profiling', for all the benchmarks. Each bargraph in this figure represents the Geometrical Mean (GMEAN) of the execution time for the number of running workloads for each category. The detailed performance of the prediction algorithms for individual workloads is presented in Figures~\ref{fig:hpc_results}--\ref{fig:parsec_results}. These figures  again confirm that a static page closure policy cannot deliver the optimum execution time for all the workloads. The corresponding workloads for HPC and SPEC benchmarks mostly prefer open-page policy. On the other hand, the corresponding workloads for PARSEC, BIOBENCH and COMMERCIAL workloads mostly prefer open-page policy.

%We have 18, 30, 26 and 26 workloads mixes for single-thread, 4-thread, 8-thread and 16-thread applications respectively. The single-thread and 4-thread workloads were run with a 1-channel memory organization while the 8-thread and 16-thread workloads are run with the 4-channel memory organization. Our experimental results show that the effect of increasing the number of ranks on prediction accuracy of predictors has the same effect as increasing the number of channels. This is because in both scenarios the memory references will be distributed between different banks which reduces the number of page conflicts. Thus, we chose to present the results with multi-channels memory organization in the paper. The detailed performance of the prediction algorithms for individual workloads is presented in Figures~\ref{fig:workload_execution_time_1_thread}--\ref{fig:workload_execution_time_16_thread}.

{\bf Overview:} Our experimental results show that the best page closure prediction scheme (i.e.\ Intel-adaptive-HAPPY) delivers 5\% and 8\% better average performance across all the workloads (up to 12\% and 20\%) in comparison with open-page and close-page policy respectively. Overall, the HAPPY implementation of both Hybrid and Intel-adaptive achieved similar performance when compared with the original implementation of these page closure policies albeit with a much lower hardware overhead. Comparing the Intel-adaptive with the Intel-adaptive-HAPPY page policy shows that the HAPPY implementation can reduce the cost of implementation by 5$\times$ for the evaluated 64~GB memory size (up to 40$\times$ for a memory size of 512~GB) while improving the performance up to 2\%. Similar behavior can be observed for Hybrid and Hybrid-HAPPY. Hybrid-HAPPY shows 182,000$\times$ reduction in cost of implementation for the evaluated 64~GB memory size (up to 1.2M$\times$ for memory size of 512~GB) while improving the performance up to 3\% for some workloads.

Similarly, as Figure~\ref{fig:multithread_thread_results} presents, for multi-thread applications, even with a very high MF, HAPPY performance is consistent and it delivers similar performance to the original implementation. The experimental results show that Intel-adaptive-HAPPY delivers 5\% and 14\% better average performance across all the workloads (up to 9\% and 22\%) in comparison with open-page and close-page policy respectively. 

%This figure also shows that although Hybrid-HAPPY performs as good as original Hybrid implementation but in general Hybrid page policy is not efficient for multithread application. The reason is that increasing the number of threads increases the randomness of memory access pattern and in this situation the page policy with the global nature (Intel-adaptive) have a faster training rate than the page policies with the 

%From Figure~\ref{fig:Final_Results}, there is a potential to improve the average performance of memory system by 7\% and 16\% (with upper bounds of 18\% and 62\%) for static open-page and close-page policies, respectively. 

\begin{figure}[!htb]
\centering
\includegraphics[scale=0.21]{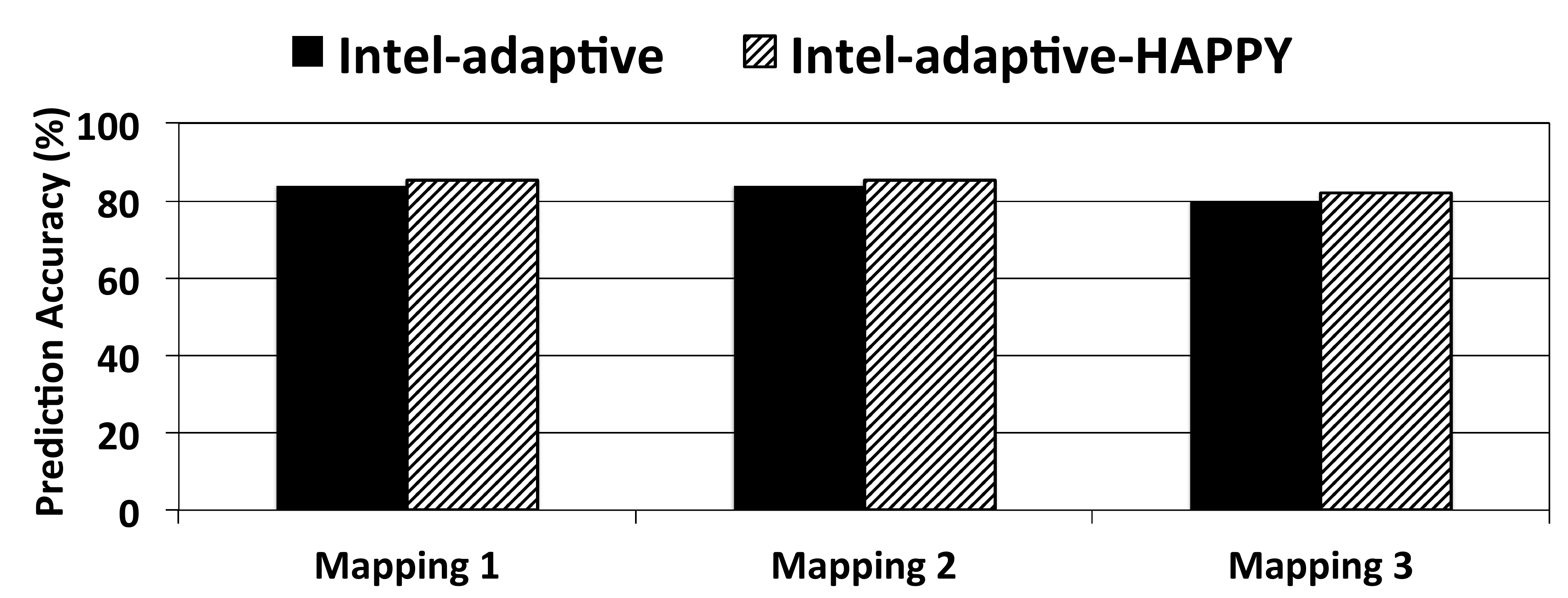}
\caption{Page-hit prediction accuracy with different address mappings.}
\label{fig:Mapping_Sens_Hit_graph}
\end{figure}

\begin{figure}[!htb]
\centering
\includegraphics[scale=0.21]{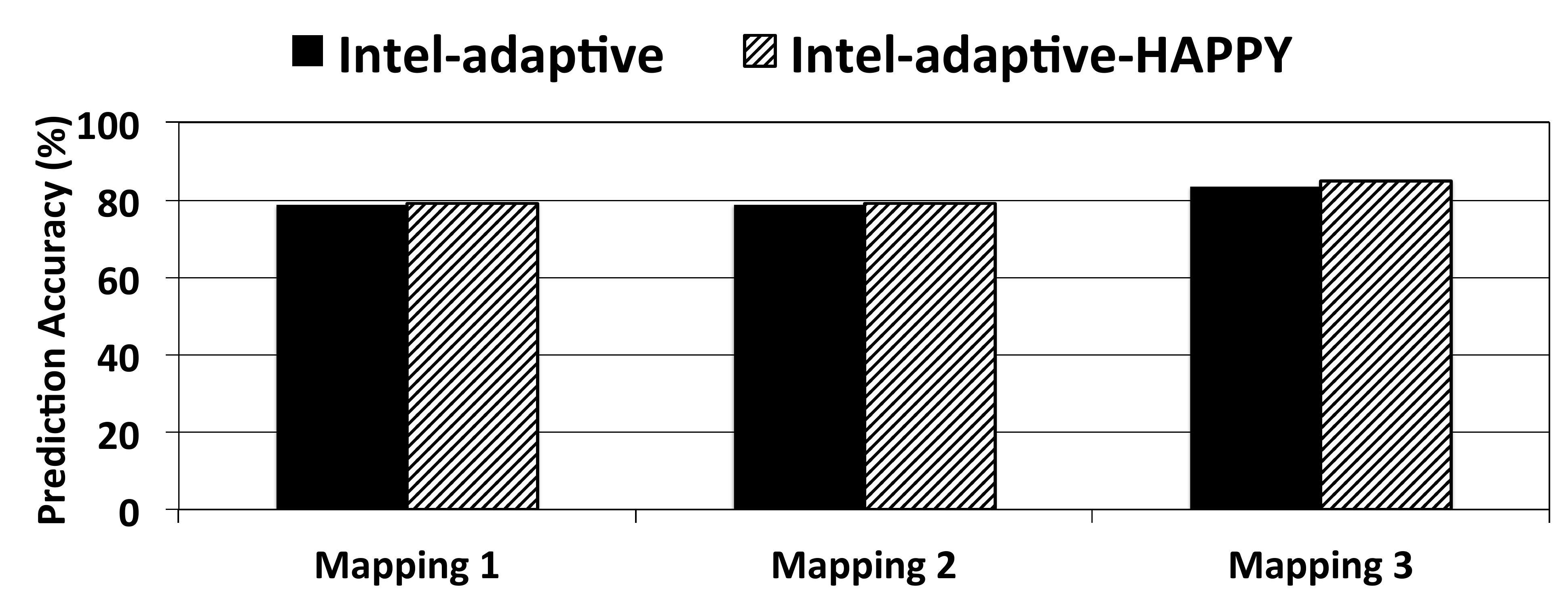}
\caption{Page-miss prediction accuracy with different address mappings.}
\label{fig:Mapping_Sens_Miss_graph}
\end{figure}

{\bf Sensitivity to Address Mapping Schemes:} To investigate the sensitivity of HAPPY to different address mappings we select the best page closure policy (i.e.\ Intel-adaptive-HAPPY) across all the predictors presented in this paper and evaluate it with the three address mappings presented in Figure~\ref{fig:Address_Mapping_fig}. Figures~\ref{fig:Mapping_Sens_Hit_graph} and  \ref{fig:Mapping_Sens_Miss_graph} illustrate the prediction accuracy of Intel-adaptive (original and HAPPY implementation) using the different mapping schemes. These results show that the HAPPY implementation of Intel-adaptive always delivers identical or slightly better results than the original implementation no matter which address mapping is used.

\begin{table*}[!htb]
\centering
  \begin{tabular}{| m{4cm} | m{12cm} | }
	\hline
	\multicolumn{2}{|c|}{\textbf{Multi-Core Workloads}} \\		
	\hline
	MIX1:  (w-D3-D3-F) & MIX12: (b-n-s-w)\\
	\hline
	MIX2:  (D1-D5-E-F) & MIX13: (w-D1-D5-x-y-z-t-F) \\
	\hline
	MIX3: (D1-x-y-F) & MIX14: (D1-D4-D5-x-z-x-B-F) \\
	\hline
	MIX4: (D2-D4-A-F) & MIX15: (D1-D4-D5-j-E-D5-v-F) \\
	\hline
	MIX5: (D2-D4-j-E) & MIX16: (D2-D4-D5-z-A-A-D4-F) \\
	\hline
	MIX6: (D2-y-t-F) & MIX17: (C5-C6-u-l-e-o-p-h) \\
	\hline
	MIX7: (D4-D5-g-F) & MIX18: (C13-C14C17-C18-C21-C2-C4-v-k-l-c-m-e-n-h-s) \\
	\hline
	MIX8: (D4-x-x-F) & MIX19: (C13-C18-C21-C2-C6-u-v-C21-u-l-l-o-t-p-h-s) \\
	\hline
	MIX9: (x-y-z-F) & MIX20: (C14-C17-C21-C22-C2-C4-C5-C8-C14-C21-C4-k-e-o-a-p) \\
	\hline
	MIX10: (C21-C22-C4-b) & MIX21: (C17-C21-u-C17-q-q-i-t-o-b-o-a-t-p-q-i) \\
	\hline
	MIX11: (C5-C6-u-e) & MIX22: (C18-C22-C5-C6-C8-u-k-l-d-e-n--o-p-q-h-r) \\	
	\hline

  \end{tabular}
  \caption{Randomly generated Multi-Core workloads. }
  \label{table:multicore_workloads}
\end{table*}

\subsection{Sensitivity to Memory Size}

We have evaluated HAPPY for up to 64~GB DRAM size and the results shows that HAPPY has a consistent behaviour as the memory size increase. Our experimental results suggests that the effective factor on HAPPY performance is the utilization of memory address space rather than the size of memory. For this reason, we considered a 4~GB memory organization with up to 99.8\% memory space utilization for our multithread experiments (results are presented in Figure~\ref{fig:multithread_thread_results}). Even in this situation, the results show that HAPPY delivers a competitive performance against the original implementation of both Hybrid and Intel-adaptive page policies while reduces the hardware overhead significantly. 

%(delivers a similar results to Figure~\ref{fig:hpc_results}-\ref{fig:multithread_thread_results})

\subsection{Scalability with Memory Size}

%It has been discussed that a main challenge in designing the page closure policy predictors is scalability. 

\begin{figure}[!htb]
\centering
\includegraphics[scale=0.18]{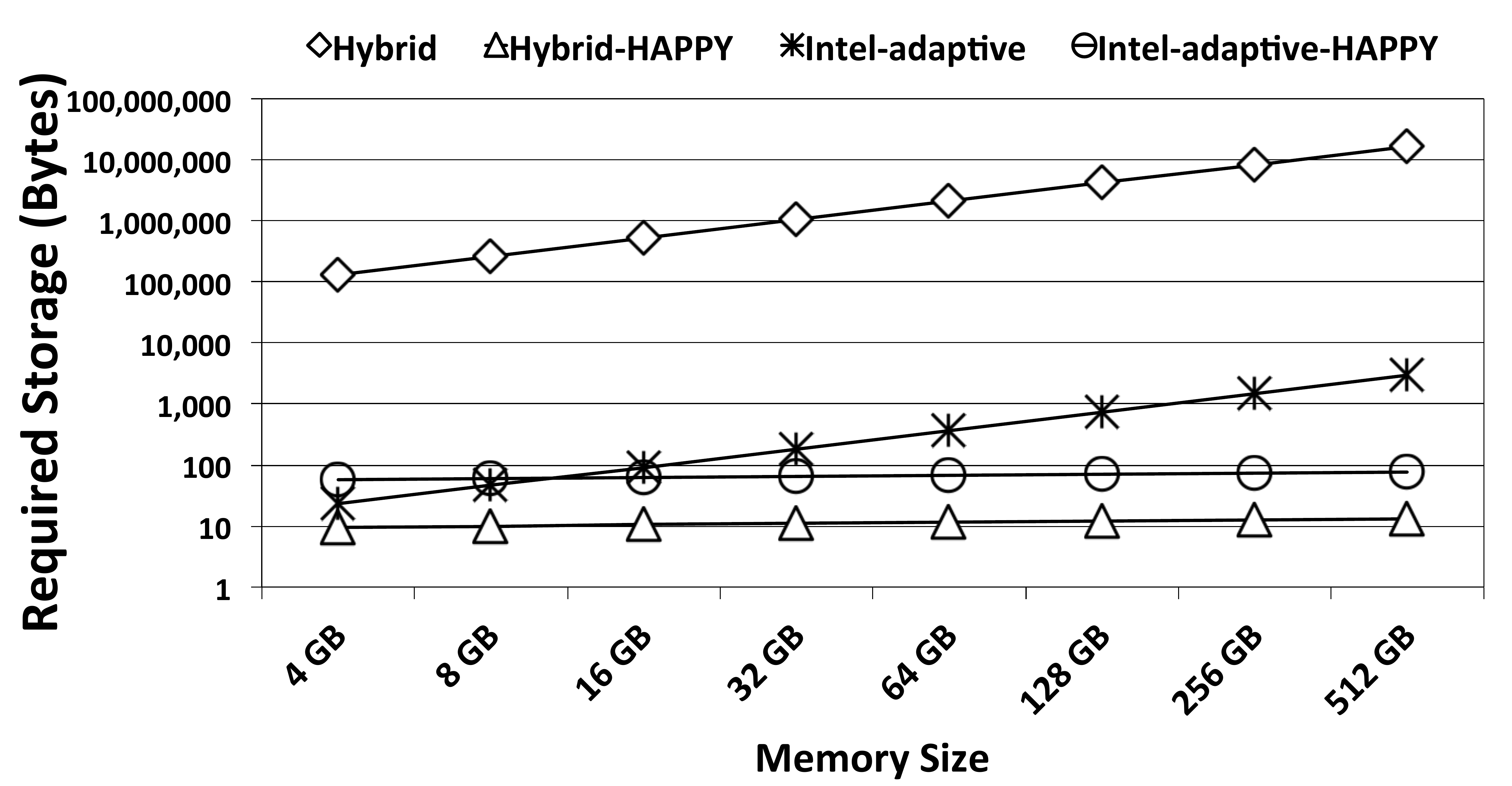}
\caption{Scalability of different page closure prediction algorithms.}
\label{fig:Scalability_graph}
\end{figure}

Figure~\ref{fig:Scalability_graph} depicts the required storage (bytes) for each prediction algorithm for different sizes of memory. The HAPPY implementation of the hybrid prediction technique is orders of magnitude (e.g.\ up to 1.2M$\times$) cheaper than the original implementation while we show that it delivers similar performance to original implementation. In the case of Intel-adaptive page closure policy, the HAPPY implementation requires slightly more resources than the original implementation for memory sizes of less than 8~GB. However, as the memory size grows, the Intel-adaptive-HAPPY outperforms the scalability over the original implementation up to 40$\times$ for a memory size of 512~GB. Table~\ref{table:Required_Counters_For_Each_Scheme} depicts the required performance counters for different page closure policies with and without HAPPY considering a memory system with X channels, Y ranks, Z banks and W rows.

\begin{table}[!htb]
\small
\centering
  \begin{tabular}{| c | c |}
	\hline
	\textbf{Implementation} & \textbf{Required Counters} \\
	\hline
	Hybrid & \(\displaystyle X\times{Y}\times{Z}\times{W} \) \\
	\hline	
	Hybrid-HAPPY & \(\displaystyle (\log_{2} X + \log_{2} Y +\log_{2} Z +\log_{2} W)\times 2 \) \\
	\hline
	Intel-adaptive & \(\displaystyle (X\times{Y}\times{Z})\times 2 \) \\
	\hline
	Intel-HAPPY & \(\displaystyle (\log_{2} X + \log_{2} Y +\log_{2} Z +\log_{2} W)\times 4 \) \\
	\hline		
  \end{tabular}
  \caption{Required performance counters for different page closure policies.}
  \label{table:Required_Counters_For_Each_Scheme}
\end{table}

%=====

\subsection{Prediction Algorithms - Weakness \& Strength}

Due to the nature of implementation and structure of each predictor, each of them might or might not work in a specific situation. Here, we discuss such situations.

{\bf Static Policies:} the open-page policy works best for high locality workloads but degrades the performance of DRAMs significantly for the workloads with highly random  or dynamic memory accesses. The close-page policy has the completely opposite behavior. PARSEC and SPEC workloads are good examples which show the different behavior of open-page and close-page policy.

%{\bf Workload-Partitioning:} this type of prediction algorithms rely on the consistency of the application behavior. They monitor memory access patterns during an observation window and apply the predicted page closure policy to future memory requests. Therefore, if the application behavior is consistent then a performance improvement can be achieved. However if the application behavior changes from one observation window to another then, for the period of the next observation window, an incorrect page closure policy will be employed that degrades the performance of the memory system. For most of presented workloads in this paper this policy does not perform well. However, for workloads like `mummer', `tigr` and `canneal` this prediction scheme performs better than Open-page policy.

{\bf Fixed-Open:} The performance of this type of algorithms is fairly susceptible to its predefined timeout value. Similarly to the methodology presented in \cite{kaseridis2011minimalist}, we select this value to be equal to t\textsubscript{RC}, that is the minimum time limitation between consecutive accesses to different rows within a bank, in the experiments. This time delay provides enough opportunity to capture a possible page hit; it does, according to the results presented in Figure~\ref{fig:Final_Results}. However, for non-memory intensive threads with high locality or memory intensive with low locality (e.g.\ `mummer' and `tigr') this technique might not work well. The reason is that, for the first category, the time interval between memory requests might be higher than the fixed timeout value which means this technique will close the row before a page hit happens. Similarly, for the second category the time interval between memory requests might be lower than the fixed timeout value, which means that a row would be kept open for an unnecessary time which, most likely, would lead to get other page conflicts.

{\bf Hybrid:} the integrated saturating counters employed in this category (either the original or HAPPY implementation) train by the number of page-hits and page-conflicts that they face. Therefore, the prediction accuracy of these types of techniques is fairly sensitive to the distribution of page hits/conflicts within DRAMs. For instance, `streamcluster'  presents such a behavior.  

{\bf Intel-adaptive:} our experiments show that this prediction algorithm is the best across all the presented schemes in this paper. However, one weakness of this technique is the updating granularity of TR. In our experiments, every time that checking of the MC suggests a more or less aggressive page closure policy, TR is incremented or decremented by one respectively.  Updating granularity by one step (increment or decrement) delivers a fine tuning of the TR but reduces the training rate of the overall prediction technique. This means that, for workloads where the application access pattern behavior changes frequently (e.g.\ between high and low locality accesses pattern) within different time phases, the Intel-adaptive scheme might not be able deliver its best performance.  Similar behavior can be observed in `canneal' or `comm1'.

{\bf HAPPY:} so far we have just explained that advantages of HAPPY. However, like all the other proposed techniques, HAPPY also has weaknesses. Considering the global nature of a HAPPY implementation it is expected that HAPPY cannot perform as efficiently as fine grain schemes for workloads with fairly dynamic behavior targeting a small part of DRAMs locally. This can be seen in workloads like `tigr'.

\subsection{Flexibility}

HAPPY is the proof of the concept that the physical address bits can be the source of useful information that can be extracted using the right encoding and decoding techniques. This makes HAPPY a fairly flexible tool that can be applied to different prediction algorithms that have not been practical due to the cost of implementation, making them feasible. In this paper we applied HAPPY to two completely different prediction schemes and showed how the performance and scalability of these scheme improved. However, page closure policies are not the only candidate that can take advantage of the knowledge presented by HAPPY in this paper and we will present more interesting show cases in the near future.

\section{Related Work}
\label{sec:Related_Work}

Succinctly, prior research in this area can be categorized in two main groups: access-based and time-based techniques. 

{\bf Access-based techniques} are those that monitor and keep a history of the row hits and row misses at different granularity in DRAMs to make a prediction of the future page closure policy for each row or bank within a DRAM memory system. Xu \textit{et al}.\ \cite{xu2009prediction} proposed a two-level dynamic SDRAM policy predictor which collects the row hit/miss behavior for the last $n$ accesses in a history register. For each entry in the history register, there is a 2-bit saturating counter that keeps track of the page closure policy for each access. Huan \textit{et al}.\ \cite{huan2006processor} proposed the Processor-Directed dynamic page policy where the processor keeps track of the last row access to each bank to predict page hits or misses for future memory requests. The processor sends this information to the memory controller to specify the page closure for the next memory access. Awashti \textit{et al}.\ \cite{awasthi2011prediction} keep track in a history table of the number of accesses each row has before closing it. When a row is open the number of expected accesses to that row is looked up, if there is no recorded entry for the accessed row in the history table that row is kept open. However if there is an entry for the accessed row it will be closed after the expected number of accesses suggested by the history table. More techniques using access-based page closure prediction can be found in \cite{park2003history,ma2007dram,miura2001dynamic,stankovic2004access,stankovic2005dram,stankovic2005dram2,schumann1997design}.

{\bf Time-based techniques} mainly focus on predicting the optimum time that a row can be left open. Blackmore \cite{blackmore2013quantitative} presented a quantitative analysis of page closure predictors. This work specifically focused on the Intel-adaptive page policy structure and tried to improve it by introducing the inter-arrival distribution concept. Stonkovic \textit{et al}.\ \cite{stankovic2005dram} used the concept of \textit{live-time} and \textit{dead-time} to predict the page closure. The live-time is the time interval between opening a row until the last access to that row while dead-time is the interval from the last access to an open row until its closing. If the predictor predicts a zero live-time or if it predicts that the row has entered its dead-time period, then the row will be closed immediately after the DRAM access otherwise it will be kept open for future accesses. In another work, Kaseridis \textit{et al}.\ \cite{kaseridis2011minimalist} used the concept that in DRAMs there is a minimum time limitation of t\textsubscript{RC} between two activations within a bank and speculatively leave the pages open for the t\textsubscript{RC} period. 

%{\bf Workload partitioning techniques} investigate the workload behavior in different time intervals and associate a page closure policy for each period. In general two main parameters are monitored during this process: workload intensity and locality. Xie \textit{et al}.\ \cite{xie2013page} classified workloads into three groups based on the workload intensity and locality: non-memory intensive, memory intensive workloads with high row-buffer locality, and memory intensive workloads with low row buffer locality. Then they assigned the close-page policy to first category to save power, the close-page policy to the second category to reduce the memory latency, and the open-page policy to the last category to exploit the row buffer locality and reduce memory latency. Kumar \textit{et al}.\ \cite{kumar2010energy} also proposed a similar technique to predict the page closure policy in DRAMs.

To sum up, HAPPY is the only technique which considers an encoding based on the memory address bits offering a compact means of storing history to inform the predictions. In addition, we have shown how to apply HAPPY to Time-based (i.e.\ Intel Adaptive HAPPY) and Access-based techniques (i.e. Hybrid HAPPY).

\section{Conclusions}

DRAM performance is dependent on the memory access pattern and, more specifically, the number of page-hit and page-conflicts that occur at run time. Modern DRAM controllers employ advanced page closure policy predictors to improve performance by trying to transform page-conflicts into page-empty (e.g.\ by closing the last accessed row at the ``right time''), and page-empty cases into page-hits (e.g.\ by keeping open the last accessed row for longer time). However the main challenge is to balance the prediction accuracy of these predictors with manageable hardware overheads (scalability) as we increase the size of DRAM.

We have described HAPPY -- a compact and efficient binary-encoding technique -- to alleviate the scalability problem of DRAM page closure predictors. HAPPY relies on the simple observation that there is a strong correlation between the physical address bits of memory addresses requested by processors and the internal structure of the DRAM as there is a fixed-address mapping scheme. Thus, the physical address bits carry the information that a memory controller needs to predict the page-hit or page-conflict for a particular access. Considering this, the required performance counters and monitoring units needed by the page closure prediction algorithms can be encoded from the physical address bits.  Doubling the size of DRAM only implies one extra physical address bit. This means that with HAPPY only one extra monitoring unit is required to predict the DRAM page closure policy when the size of memory is doubled. In other words, HAPPY offers the smallest hardware overhead to implement dynamic DRAM page closure predictor algorithms.

We have evaluated HAPPY by applying it to a traditional Hybrid page closure policy, as well as the state-of-the-art Intel-adaptive open page policy included in Intel Xeon X5650. The experimental results show that the HAPPY implementation of Intel-adaptive page policy can reduce the cost of implementation by 5$\times$ for the evaluated 64~GB memory size (up to 40$\times$ for a memory size of 512~GB) while maintaining the prediction accuracy. The other case study shows 182,000$\times$ reduction in cost of implementation for the evaluated 64~GB memory size (up to 1.2M$\times$ for memory size of 512~GB) while maintaining the prediction accuracy. The experiments have also reported the accuracy of the predictors and have studied the sensitivity towards the memory address-mapping. In both scenarios, HAPPY maintains its key advantage of offering no degradation of prediction accuracy while reducing significantly the hardware overhead.

%To summarize, considering the basic principles behind HAPPY, it can be applied not only to page closure prediction algorithms but also to any other similar computational algorithm in DRAMs to improve the scalability and exploiting the encoded information in memory access pattern at run time.

%%%%%%% -- PAPER CONTENT ENDS -- %%%%%%%%

\section{Acknowledgements}
The research leading to these results has received funding from the European Union's Seventh Framework Programme (FP7/2007-2013) under grant agreement n$^{\circ}$ 318633; AXLE project http://axleproject.eu/. Mikel Luj{\'a}n is funded by a Royal Society University Research Fellowship and further supported by UK EPSRC grants DOME EP/J016330/1 and PAMELA EP/K008730/1.

%%%%%%%%% -- BIB STYLE AND FILE -- %%%%%%%%
\bibliographystyle{ieeetr}
\bibliography{references}

\begin{thebibliography}{10}

\bibitem{website:RAMCloud}
{ John Ousterhout}, ``Ramcloud.''

\bibitem{ongaro2011fast}
D.~Ongaro, S.~M. Rumble, R.~Stutsman, J.~Ousterhout, and M.~Rosenblum, ``Fast
  crash recovery in ramcloud,'' in {\em Proceedings of the Twenty-Third ACM
  Symposium on Operating Systems Principles}, pp.~29--41, ACM, 2011.

\bibitem{jacob2010memory}
B.~Jacob, S.~Ng, and D.~Wang, {\em Memory systems: cache, DRAM, disk}.
\newblock Morgan Kaufmann, 2010.

\bibitem{itoh2001vlsi}
K.~Itoh, {\em VLSI memory chip design}, vol.~5.
\newblock Springer New York, 2001.

\bibitem{keeth2008dram}
B.~Keeth, {\em DRAM circuit design: fundamental and high-speed topics},
  vol.~13.
\newblock Wiley. com, 2008.

\bibitem{ousterhout2010case}
J.~Ousterhout, P.~Agrawal, D.~Erickson, C.~Kozyrakis, J.~Leverich,
  D.~Mazi{\`e}res, S.~Mitra, A.~Narayanan, G.~Parulkar, M.~Rosenblum, {\em
  et~al.}, ``The case for ramclouds: scalable high-performance storage entirely
  in dram,'' {\em ACM SIGOPS Operating Systems Review}, vol.~43, no.~4,
  pp.~92--105, 2010.

\bibitem{dodd2006adaptive}
J.~Dodd, ``Adaptive page management,'' July~11 2006.
\newblock US Patent 7,076,617.

\bibitem{website:inteladaptive}
{Rajinder Gill}, ``Everything you always wanted to know about sdram memory but
  were afraid to ask.''

\bibitem{Intelweb}
Intel, ``Intel xeon processor x5650.''

\bibitem{brown2010}
G.~Brown, ``Ensemble learning,'' {\em Encyclopedia of Machine Learning}, 2010.

\bibitem{chatterjee2012usimm}
N.~Chatterjee, R.~Balasubramonian, M.~Shevgoor, S.~Pugsley, A.~Udipi,
  A.~Shafiee, K.~Sudan, M.~Awasthi, and Z.~Chishti, ``Usimm: the utah simulated
  memory module,'' {\em University of Utah, Tech. Rep}, 2012.

\bibitem{kaseridis2011minimalist}
D.~Kaseridis, J.~Stuecheli, and L.~K. John, ``Minimalist open-page: A dram
  page-mode scheduling policy for the many-core era,'' in {\em Proceedings of
  the 44th Annual IEEE/ACM International Symposium on Microarchitecture},
  pp.~24--35, ACM, 2011.

\bibitem{zhang2000permutation}
Z.~Zhang, Z.~Zhu, and X.~Zhang, ``A permutation-based page interleaving scheme
  to reduce row-buffer conflicts and exploit data locality,'' in {\em
  Proceedings of the 33rd annual ACM/IEEE international symposium on
  Microarchitecture}, pp.~32--41, ACM, 2000.

\bibitem{bienia2008parsec}
C.~Bienia, S.~Kumar, J.~P. Singh, and K.~Li, ``The parsec benchmark suite:
  characterization and architectural implications,'' in {\em Proceedings of the
  17th international conference on Parallel architectures and compilation
  techniques}, pp.~72--81, ACM, 2008.

\bibitem{dixit1991spec}
K.~M. Dixit, ``The spec benchmarks,'' {\em Parallel computing}, vol.~17,
  no.~10, pp.~1195--1209, 1991.

\bibitem{albayraktaroglu2005biobench}
K.~Albayraktaroglu, A.~Jaleel, X.~Wu, M.~Franklin, B.~Jacob, C.-W. Tseng, and
  D.~Yeung, ``Biobench: A benchmark suite of bioinformatics applications,'' in
  {\em Performance Analysis of Systems and Software, 2005. ISPASS 2005. IEEE
  International Symposium on}, pp.~2--9, IEEE, 2005.

\bibitem{xu2009prediction}
Y.~Xu, A.~S. Agarwal, and B.~T. Davis, ``Prediction in dynamic sdram controller
  policies,'' in {\em Embedded Computer Systems: Architectures, Modeling, and
  Simulation}, pp.~128--138, Springer, 2009.

\bibitem{huan2006processor}
D.~Huan, Z.~Li, W.~Hu, and Z.~Liu, ``Processor directed dynamic page policy,''
  in {\em Advances in Computer Systems Architecture}, pp.~109--122, Springer,
  2006.

\bibitem{awasthi2011prediction}
M.~Awasthi, D.~W. Nellans, R.~Balasubramonian, and A.~Davis, ``Prediction based
  dram row-buffer management in the many-core era,'' in {\em Parallel
  Architectures and Compilation Techniques (PACT), 2011 International
  Conference on}, pp.~183--184, IEEE, 2011.

\bibitem{park2003history}
S.-I. Park and I.-C. Park, ``History-based memory mode prediction for improving
  memory performance,'' in {\em Circuits and Systems, 2003. ISCAS'03.
  Proceedings of the 2003 International Symposium on}, vol.~5, pp.~V--185,
  IEEE, 2003.

\bibitem{ma2007dram}
C.~Ma and S.~Chen, ``A dram precharge policy based on address analysis,'' in
  {\em Digital System Design Architectures, Methods and Tools, 2007. DSD 2007.
  10th Euromicro Conference on}, pp.~244--248, IEEE, 2007.

\bibitem{miura2001dynamic}
S.~Miura, K.~Ayukawa, and T.~Watanabe, ``A dynamic-sdram-mode-control scheme
  for low-power systems with a 32-bit risc cpu,'' in {\em Proceedings of the
  2001 international symposium on Low power electronics and design},
  pp.~358--363, ACM, 2001.

\bibitem{stankovic2004access}
V.~Stankovi{\'c} and N.~Milenkovi{\'c}, ``Access latency reduction in
  contemporary dram memories,'' {\em Facta universitatis-series: Electronics
  and Energetics}, vol.~17, no.~1, pp.~81--97, 2004.

\bibitem{stankovic2005dram}
V.~V. Stankovic and N.~Z. Milenkovic, ``Dram controller with a close-page
  predictor,'' in {\em Computer as a Tool, 2005. EUROCON 2005. The
  International Conference on}, vol.~1, pp.~693--696, IEEE, 2005.

\bibitem{stankovic2005dram2}
V.~Stankovic and N.~Milenkovic, ``Dram controller with a complete predictor:
  Preliminary results,'' in {\em Telecommunications in Modern Satellite, Cable
  and Broadcasting Services, 2005. 7th International Conference on}, vol.~2,
  pp.~593--596, IEEE, 2005.

\bibitem{schumann1997design}
R.~C. Schumann, ``Design of the 21174 memory controller for digital personal
  workstations,'' {\em Digital Technical Journal}, vol.~9, pp.~57--70, 1997.

\bibitem{blackmore2013quantitative}
M.~Blackmore, ``A quantitative analysis of memory controller page policies,''
  {\em Notes}, vol.~2013, pp.~01--01, 2013.

\end{thebibliography}
%%%%%%%%%%%%%%%%%%%%%%%%%%%%%%%%%%%%

%\input{Sections/Appendix}

\end{document}